\pdfoutput=1

\documentclass[sigconf]{acmart} 
\usepackage{booktabs}
\usepackage{multirow}
\newcommand*\rot{\rotatebox{90}}
\usepackage{xspace}
\usepackage{subcaption}
\usepackage{amsmath}
\usepackage{enumitem}
\usepackage{tabularx}
\usepackage{balance}
\usepackage{color, colortbl}
\usepackage{boldline}
\usepackage[ruled,vlined,linesnumbered]{algorithm2e}
\usepackage[compatible]{algpseudocode}
\renewcommand{\algorithmiccomment}[1]{\bgroup\hfill\small//~#1\egroup}
\usepackage{makecell}

\newlength{\textfloatsepsave}  

\newlist{researchquestions}{enumerate}{1}
\setlist[researchquestions]{label*=\textbf{RQ\arabic*}}

\usepackage{cleveref}

\crefformat{section}{\S#2#1#3}
\crefformat{subsection}{\S#2#1#3}
\crefformat{subsubsection}{\S#2#1#3}

\AtBeginDocument{%
 }

\copyrightyear{2024}
\acmYear{2024}
\setcopyright{rightsretained}
\acmConference[WWW '24]{Proceedings of the ACM Web Conference 2024}{May 13--17, 2024}{Singapore, Singapore}
\acmBooktitle{Proceedings of the ACM Web Conference 2024 (WWW '24), May 13--17, 2024, Singapore, Singapore}\acmDOI{10.1145/3589334.3645512}
\acmISBN{979-8-4007-0171-9/24/05}

\makeatletter
\gdef\@copyrightpermission{
  \begin{minipage}{0.3\columnwidth}
   \href{https://creativecommons.org/licenses/by/4.0/}{\includegraphics[width=0.90\textwidth]{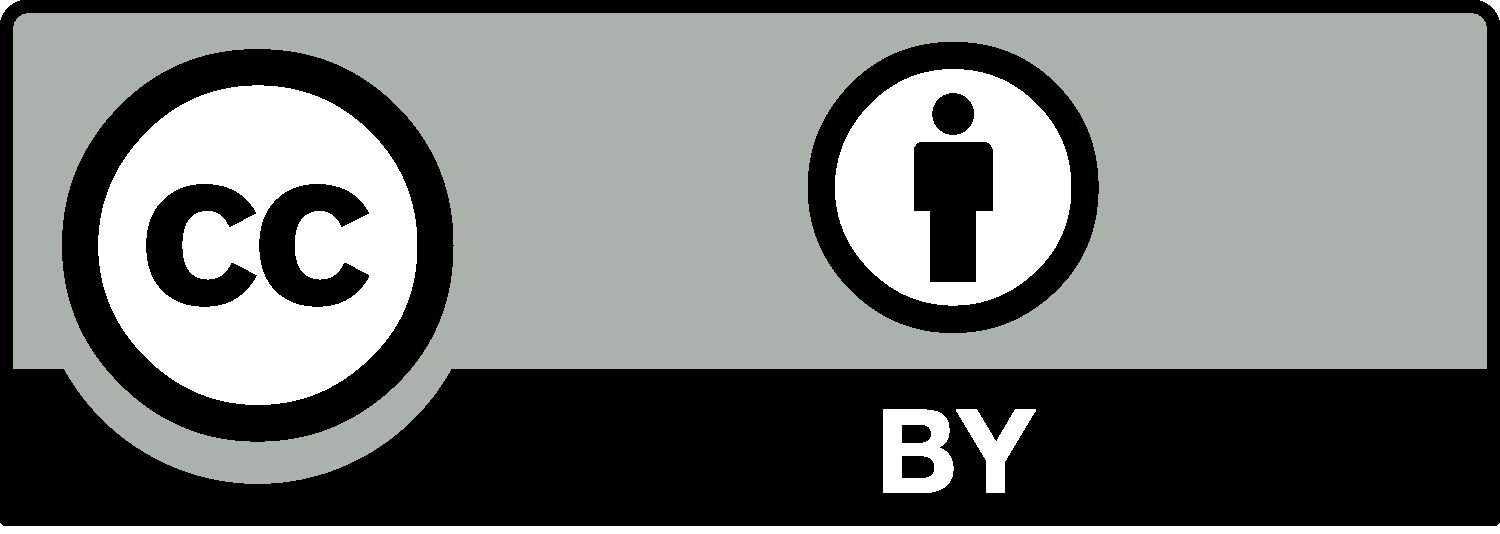}} %{\includegraphics[width=0.90\textwidth]{figure/4ACM-CC-by-88x31.eps}}
  \end{minipage}\hfill
  \begin{minipage}{0.7\columnwidth}
   \href{https://creativecommons.org/licenses/by/4.0/}{This work is licensed under a Creative Commons Attribution International 4.0 License.}
  \end{minipage}
  \vspace{5pt}
}
\makeatother

\newcommand{\proposed}{ToTER\xspace}

\newcommand{\smallsection}[1]{{\vspace{0.02in} \noindent \bf {#1.}}}

\newcommand{\CTR}{Contriever\xspace}
\newcommand{\QE}{BQE\xspace}
\newcommand{\TQE}{TopicGQA\xspace}

\begin{document}

\title{Improving Retrieval in Theme-specific Applications \\using a Corpus Topical Taxonomy}

\author{SeongKu Kang}
\affiliation{
    \institution{University of Illinois at Urbana-Champaign}
    \city{IL}
    \country{USA}
}
\email{seongku@illinois.edu}

\author{Shivam Agarwal}
\affiliation{
    \institution{University of Illinois at Urbana-Champaign}
    \city{IL}
    \country{USA}
}
\email{shivama2@illinois.edu}

\author{Bowen Jin}
\affiliation{
    \institution{University of Illinois at Urbana-Champaign}
    \city{IL}
    \country{USA}
}
\email{bowenj4@illinois.edu}

\author{Dongha Lee}
\affiliation{
    \institution{Yonsei University}
    \city{Seoul}
    \country{Republic of Korea}
}
\email{donalee@yonsei.ac.kr}

\author{Hwanjo Yu}
\affiliation{
    \institution{Pohang University of\\ Science and Technology}
    \city{Pohang}
    \country{Republic of Korea}
}
\authornote{Corresponding author.}
\email{hwanjoyu@postech.ac.kr}

\author{Jiawei Han}
\affiliation{
    \institution{University of Illinois at Urbana-Champaign}
    \city{IL}
    \country{USA}
}
\email{hanj@illinois.edu}

\renewcommand{\shortauthors}{SeongKu Kang et al.}
\begin{abstract}
  Document retrieval has greatly benefited from the advancements of large-scale pre-trained language models (PLMs). % owing to their superior capability to understand contextual semantics.
However, their effectiveness is often limited in theme-specific applications for specialized areas or industries, due to unique terminologies, incomplete contexts of user queries, and specialized search intents.
To capture the theme-specific information and improve retrieval, we propose to use a \textit{corpus topical taxonomy}, which outlines the latent topic structure of the corpus while reflecting user-interested aspects. 
We introduce \proposed (Topical Taxonomy Enhanced Retrieval) framework, which identifies the central topics of queries and documents with the guidance of the taxonomy, and exploits their topical relatedness to supplement missing contexts.
As a plug-and-play framework, \proposed can be flexibly employed to enhance various PLM-based retrievers.
Through extensive quantitative, ablative, and exploratory experiments on two real-world datasets, we ascertain the benefits of using topical taxonomy for retrieval in theme-specific applications and demonstrate the effectiveness~of~\proposed.

% The constructed taxonomy can be subsequently employed to provide additional clues to link queries and documents by discerning their topical relatedness and supplementing the missing contexts.
% To harness the corpus-level taxonomy for document-level retrieval, \proposed first conducts \textit{topic class relevance learning} which unveils the relevance of each document to each topic node in the taxonomy.
% Then, \proposed introduces a series of \textit{strategies to complement the existing retrieval} pipeline, which exploits topical relatedness between queries and documents to supplement missing contexts and improve retrieval accuracy.
% We put forward \proposed as a plug-and-play framework compatible with various retrieval models, and introduce strategies to exploit topical relatedness between queries and documents.
% Our comprehensive experiments on two real-world datasets demonstrate that \proposed significantly improves the existing PLM-based retrieval in theme-specific applications. 

\end{abstract}




\maketitle

\section{Introduction}
% Pre-trained language models (PLMs) have improved document retrieval owing to their superior semantic understanding \cite{CTR, ANCE, colbert, DPR}. 
Pre-trained language models (PLMs) have greatly improved document retrieval \cite{CTR, ANCE, colbert, DPR}. 
The PLM-based retrieval models are first pre-trained on the massive textual corpora to grasp language understanding.
Subsequently, they are fine-tuned using vast datasets of annotated query-document pairs, which enables the models to capture their semantic similarities. % within the latent space. 
While successful in general domains like web search which consist of a broad user base, they are often limited in \textit{specialized applications with specific~themes}. 
% Despite their remarkable capability, their effectiveness is often limited in \textit{specialized applications with specific themes}, which have distinct natures from the general domain \cite{li2023sailer, chaudhary2023exploring}. 

Theme-specific applications are specialized areas or industries where retrieval tasks are focused on a specific theme (e.g., academic paper search, product search in e-commerce).
Retrieval in theme-specific applications poses three challenges spanning specialized terminology and niche content (C1), limited contexts of user query (C2), and specialized user interests and search intents (C3).

\vspace{0.05cm} \noindent
\textbf{C1}: Theme-specific domains often have specialized terminologies, which are not frequently included in the general text corpus.
For example, Table \ref{tab:example}(a) shows that an academic paper includes many technical terms specific to certain research fields, such as ``proof of retrievability'' and ``cryptographic proof''. 
PLM-based retrievers trained on general text corpora often lack an inherent understanding of domain-specific specialized and niche terminologies \cite{dong2022incorporating}.

\vspace{0.05cm} \noindent
\textbf{C2}: Users familiar with the domain often omit contexts they believe are naturally implied in their query.
For example, in product search, users enter a query such as ``RTX 3090'' without adding contexts such as ``graphics cards''. 
Table \ref{tab:example}(a) shows queries from domain experts may skip over general contexts such as ``cryptography'' or ``computer security''. 
Omitted terms hinder the model's ability to fully comprehend the query, leading to imprecise retrieval outcomes. 
Inferring missing contexts is more challenging in theme-specific applications as it often requires domain-specific knowledge.

% Lastly, thank you for pointing out the insufficient descriptions on the tables. We will refine the tables and add more descriptions to the caption. For Table 1, we will explicitly denote the role of each attribute, including (1) the relevance annotation: relevant/irrelevant, (2) predicted rank by retrieval, and (3) Predicted topic classes and core phrases by ToTER.

\begin{table*}[t]
\caption{Examples of retrieval in theme-specific applications.
We use \CTR-MS (retriever) and MiniLM-L-12 (reranker). 
Contents closely related to the query are denoted in bold. 
Details of topic class and core phrase discovery are provided in \cref{sec:method}.}
\large
\renewcommand{\tabcolsep}{0.9mm}
\resizebox{1.\linewidth}{!}{
\begin{tabular}{p{2.2cm}|p{8.5cm}|p{2.3cm}|p{8.5cm}}
\hlineB{2.5}
\multicolumn{2}{c|}{(a) Academic domain} & \multicolumn{2}{c}{(b) Product domain} \\ \hline\hline
Query & Provable data possession at untrusted stores & Query & \#1 black natural hair dye without ammonia or peroxide \\ \hline 
\multirow{1}{=}{Document A\\(label: relevant \\ rank: top-173)} 
& \multirow{2}{=}{\parbox{8.3cm}{\vspace{2pt}Pors: \textbf{proofs of retrievability} for large files. In this paper, we define and explore proofs of retrievability (PORs). ... A POR may be viewed as a kind of \textbf{cryptographic proof} of knowledge (POK). ... We view PORs as an important tool for \textbf{semi-trusted} online archives. Existing cryptographic techniques help users \textbf{ensure the privacy and integrity} of files they retrieve. ...}}
& \multirow{1}{=}{Document A \\ (label: relevant \\rank: top-70)}
& ONC NATURALCOLORS (1N \textbf{Black}) 4 fl. oz. (120 mL). Healthier permanent \textbf{hair dye} with certified organic ingredients, \textbf{ammonia free}, vegan friendly, 100 gray coverage. ...  Cruelty-free and vegan. It is time to make the clean choice. \\ \cline{3-4}
 &  & \multirow{1}{=}{Document B\\(label:~irrelevant \\ rank: top-11)} & Roux Fanci-full Rinse 16 Hidden Honey. Tones and enhances gray and blonde hair. Rinses in and shampoos out. \textbf{No ammonia or peroxide}. … 15 applications per bottle, temporary \textbf{hair color}, 15 ounce bottle. \\ \hlineB{2.5}
\multirow{2}{=}{\proposed rank: \\ top-10}& Topic classes: cryptography, trusted computing, digital content, computer network, computer security, computer science &  \multirow{2}{=}{\proposed rank: \\top-5 (Doc.A)\\top-32~(Doc.B)}  & Topic classes: hair color, hair coloring products, hair care, beauty \& personal care \\ \cline{2-2}\cline{4-4}
 & \multirow{1}{=}{Core phrases: encryption, access control, security, key, server} &  & Core phrases: dye, permanent, lasting, permanent hair color, ammonia free\\ 
\hlineB{2.5}
\end{tabular}}
\label{tab:example}
\vspace{-0.2cm}
\end{table*}

% Users in different applications have different search intents, and the concept of relevance can be different according to the user interests within the domain.
\vspace{0.05cm} \noindent    
\textbf{C3}: Users in theme-specific applications have more specialized interests and intents compared to general web searches. 
For example, researchers may want to find papers within a specific field of study to discern a particular research trajectory. 
In product search, users often filter results based on specific product attributes.
For example, Table \ref{tab:example}(b) shows that both documents are somewhat relevant to the query as both of them are about ammonia-free hair color products. 
However, the query targets hair dye with lasting effects, instead of hair rinse with temporary effects.
These specialized search intents are not effectively captured by models trained on general corpora.

Accumulating ample labeled data can mitigate these challenges to some extent. 
However, the creation of such datasets in theme-specific applications is particularly challenging due to the need for domain expertise (e.g., academic domain) and the proprietary nature of user logs in specialized applications (e.g., e-commerce) \cite{chaudhary2023exploring, li2023sailer}.
As a result, PLM-based retrieval models often struggle to accurately capture relevance in theme-specific applications \cite{thakur2021beir}.

% Moreover, a popular query-generation approach, which uses synthetically generated queries (from a PLM-based model) as a substitute for real labeled data, often leads to unsatisfactory outcomes in these specialized applications~\cite{chaudhary2023exploring}.

% the potential benefits of using topical taxonomy
To improve retrieval without relying on labeled data, we propose to use a \textit{corpus topical taxonomy} \cite{huang2020corel, lee2022taxocom, meng2020hierarchical, shang2020nettaxo, zhang2018taxogen}, which has been extensively studied for organizing topics in a corpus.
A corpus topical taxonomy outlines the latent topic hierarchy within the corpus as a tree structure, where each node is a \textit{topic class} represented by a cluster of semantically coherent terms describing the topic, as shown in Figure \ref{fig:method}.
Recent taxonomy construction studies \cite{huang2020corel, lee2022taxocom, arous2023taxocomplete} have effectively reflected user-interested aspects, drawing from a foundational seed taxonomy rooted in human knowledge of the application (e.g., fields of study from Mircosoft Academic \cite{MAG_FS}).
% The constructed taxonomy can be subsequently employed to reduce the reliance on extensive human labeling for downstream tasks, such as classification \cite{shen2021taxoclass} and recommendation \cite{tan2022enhancing}.
The constructed taxonomy can be subsequently employed to provide additional clues to link queries and documents by discerning their topical relatedness and supplementing the missing contexts.
% However, the potential of such taxonomies in enhancing PLM-based retrieval remains unexplored in the previous literature.

We propose \textbf{To}pical \textbf{T}axonomy \textbf{E}nhanced \textbf{R}etrieval (\proposed) framework, which systematically leverages the corpus taxonomy to complement the semantic matching of PLM-based retrieval.
The taxonomy provides a high-level topic hierarchy of the entire corpus. 
To harness this corpus-level knowledge for retrieval, we first link it to individual documents.
Specifically, \proposed first conducts \textit{topic class relevance learning} to discern the relevance of each document to each topic class node in the taxonomy.
% As it is infeasible to obtain topic annotations for all documents, 
We formulate this step as an unsupervised multi-label classification problem without document-topic labels.
\proposed introduces a new silver label generation strategy along with a new collective distillation process to produce rich and reliable signals.
This class relevance learning allows \proposed to effectively identify central subjects of a given text under the guidance of the topical taxonomy reflecting user interests.

Based on the identified topic class relevance, \proposed leverages the topical relatedness of a query and documents to complement the semantic matching by PLM-based retrievers.
In Table 1(a), we see that \proposed can improve retrieval by identifying common topic classes like ``cryptography'' and ``computer security'' for both query and document, given the presence of terms frequently used for these topic classes (\textbf{C1}).
Furthermore, \proposed combines the topical relatedness with more fine-grained phrase knowledge for each topic class, helping to distinguish documents having similar topics.
In Table 1(b), \proposed identifies and utilizes core topical phrases such as ``dye'', ``lasting'', and ``permanent hair color'' to enrich the query, enabling more accurate finding of relevant documents (\textbf{C2}).
This entire process is built upon the topical taxonomy reflecting user-interested aspects (\textbf{C3}).
Formally, \proposed introduces three strategies to complement the PLM-based retrieval: (1) search space adjustment, (2) class relevance matching, and (3) query enrichment by core phrases.
Our contributions are summarized as follows:
\begin{itemize}[leftmargin=*]\vspace{-\topsep}
    \item We present a systematic approach to incorporate topical taxonomy into retrieval in theme-specific applications, which remains unexplored in the previous literature.
    % which is new for PLM-based retrieval.
    \item We propose \proposed that deliberately discerns and utilizes topical relatedness of queries and documents.
     As a plug-and-play framework, it can be integrated with various PLM-based models.
    \item We validate the effectiveness of \proposed by extensive experiments. 
    \proposed consistently improves retrieval accuracy in scenarios with both no labeled data and limited labels.
    % Furthermore, we provide an in-depth analysis of the \proposed framework.
\end{itemize}\vspace{-\topsep}

\section{Problem Formulation}
\subsection{Concept Definition}
\label{subsec:concept}
\noindent
\textbf{PLM-based multi-stage retrieval.}
Most PLM-based retrieval systems leverage multi-stage retrieve-then-rerank pipeline \cite{ma2020zero, dong2022incorporating, AR2}. 
Specifically, given a query $q$, a \underline{\textit{retriever}} retrieves a set of candidate documents $\mathcal{D}_q$ from a large corpus $\mathcal{D}$, where $|\mathcal{D}_q|\ll|\mathcal{D}|$.
Following the first-stage retrieval, a \underline{\textit{reranker}} computes a more fine-grained relevance for each candidate document, and generates the final ranked list by reordering them.

The first-stage retriever typically adopts a dual-encoder architecture, where query and documents are encoded separately and the relevance is measured by the similarity between their embeddings. 
\begin{equation}
    s_{de}(q,d) = sim(f_{\theta}(q), f_{\theta'}(d)),
\end{equation}
where $f_\theta$ and $f_{\theta'}$ are the query and document encoders, and $sim(\cdot,\cdot)$ is the similarity function such as the cosine similarity.
The document embeddings are pre-computed and efficiently retrieved via approximate nearest-neighbor (ANN) search techniques \cite{johnson2019billion}.

The second-stage reranker mostly adopts a cross-encoder architecture which takes the concatenation of a query and a document as input and assesses its relevance score. 
\begin{equation}
    s_{ce}(q,d) = f_{\phi}(q, d),
\end{equation}
where $f_\phi$ denotes the reranker.
By fully exploring the interactions between the query and document, it generates more accurate relevance scores compared to the dual-encoder \cite{AR2}.
% $\pi_q$ is generated by reordering $\mathcal{D}_q$ based on their corresponding reranker scores.

\begin{figure*}[t]
\centering
\includegraphics[width=1.0\textwidth]{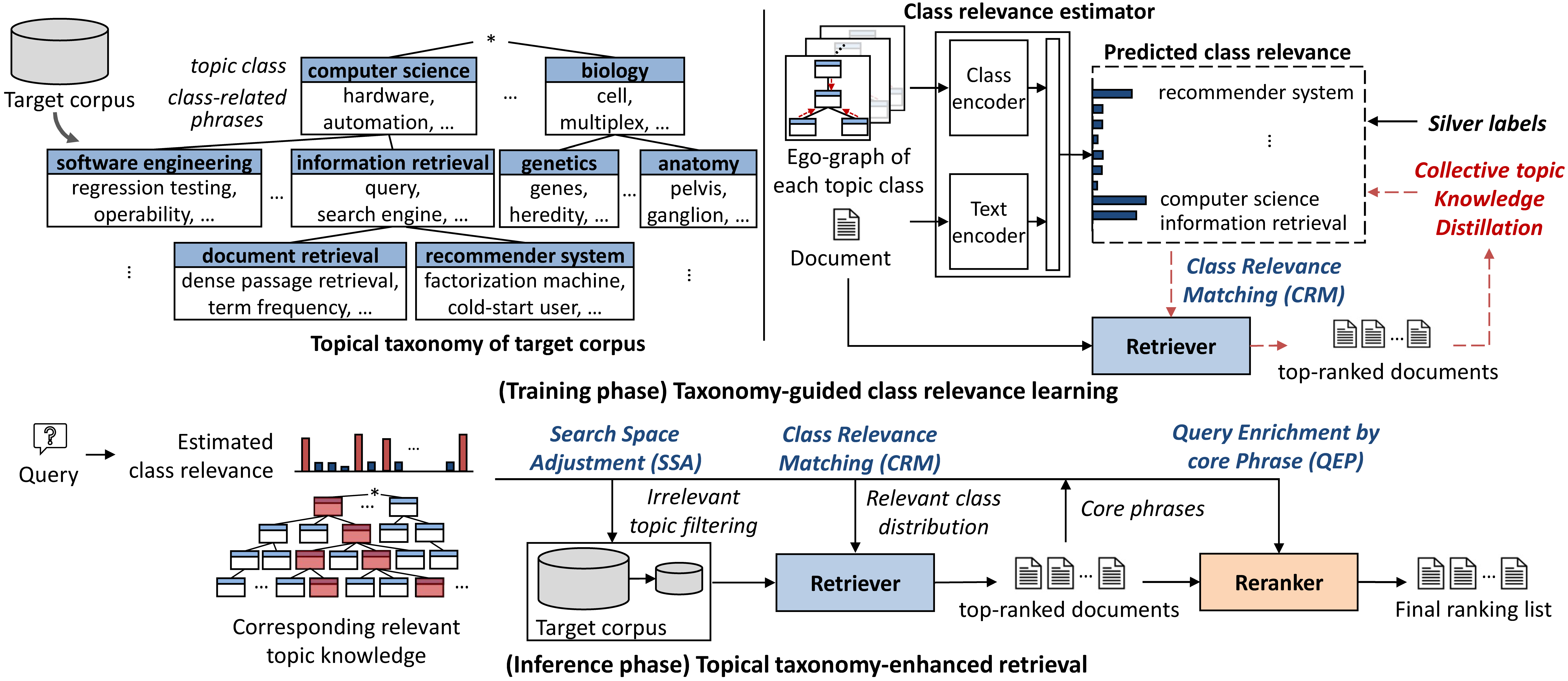}
\caption{The overview of Topical Taxonomy-Enhanced Retrieval (\proposed) framework.
}
\label{fig:method}
\vspace{-0.2cm}
\end{figure*}

\noindent
\textbf{Topical taxonomy.}
A \underline{\textit{corpus topical taxonomy}} $\mathcal{T} = (\mathcal{C}, \mathcal{R})$ represents a tree structure that outlines the latent topic hierarchy within the target corpus.
Each node $c_j \in \mathcal{C}$ corresponds to a \underline{\textit{topic class}} which is described by a coherent cluster of terms\footnote{Each term is regarded as a phrase composed of one or multiple word tokens, so the terms ``phrase'' and ``term'' are used interchangeably in this paper.} describing the topic, denoted by $P_j$.
The most salient term (i.e., center term) is utilized as the class name.
Each edge ($\in \mathcal{R}$) indicates a theme-specific relationship between a parent and child node, such as “is a subfield of” or “is a type of”.
Figure \ref{fig:method} shows an example of topical taxonomy.

To construct the topical structure reflecting theme-specific user interests, taxonomies are built upon a foundational seed taxonomy rooted in human knowledge of the application \cite{huang2020corel, lee2022taxocom, arous2023taxocomplete}.
An example is the fields of study in the academic domain \cite{MAG_FS}, which embodies researchers' inclination to organize academic concepts and studies.
Based on this application knowledge, recent methods \cite{huang2020corel, lee2022taxocom, zhang2023effective, lee2022topicgen} have effectively generated taxonomy having remarkable term coherency, topic coverage, and user-interest alignment.

\subsection{Problem Definition}
% We focus on retrieval within theme-specific applications, which are systems tailored for particular fields or industries.
We focus on retrieval in theme-specific applications, which are specialized areas or industries where retrieval tasks are centered around specific themes.
These applications differ from general web search in terms of (1) domain-specialized terminology and niche content, (2) expert users familiar with the domain, and (3) specialized search intents.
Given a target corpus $\mathcal{D}$ and its topical taxonomy $\mathcal{T}$\footnote{The topical taxonomy can be obtained using any off-the-shelf taxonomy completion technique.
In this work, we use the recently proposed method \cite{lee2022taxocom}.}, 
our goal is to develop a systematic framework that exploits the topic hierarchy knowledge to improve the existing PLM-based multi-stage retrieval. 
We focus on scenarios where labeled data from the target corpus is unavailable.
It is worth noting that we pursue a plug-and-play solution that can be flexibly integrated with various existing retriever and reranker models.

\section{Related Work}
\label{sec:relatedwork}
\smallsection{Topical taxonomy completion}
Topical taxonomy represents the latent topic hierarchy of document collections, providing valuable knowledge of contents in many applications.
Early methods build a corpus topic taxonomy from scratch by extracting discriminative term clusters from the corpus in a hierarchical fashion \cite{zhang2018taxogen}.
To generate user-interested topic structure, recent approaches \cite{shang2020nettaxo, lee2022taxocom, huang2020corel, lee2022topicgen} have started with a seed taxonomy rooted in human knowledge of the application and expanded it by discovering novel topics from the target corpus.
% The seed taxonomy is used to guide the whole process of discovering new topics and expanding the taxonomy.
% They use the initial taxonomy to supervise the learning of hierarchical relations. 
Specifically, \cite{huang2020corel} trains classifiers to capture user-interested relations from parent-child topic pairs, \cite{lee2022taxocom} recursively clusters phrases to identify new subtopics based on the known topic relations.
Very recently, \cite{lee2022topicgen} generates topic-conditioned terms by leveraging hierarchical relations from the seed taxonomy.

\smallsection{PLM-based retrieval models.}
PLM-based retrieval models have advanced in both training and encoding strategies.
In terms of training strategy, starting from in-batch and hard negatives by BM25 \cite{DPR, luan2021sparse}, advanced hard negative mining by dynamic sampling \cite{zhan2021optimizing} and denoising using a cross-encoder \cite{rocketqa_v1} have been studied.
Many works have focused on pre-training with unsupervised contrastive learning \cite{CTR, simcse, condenser}, knowledge graph \cite{kim2021query, dong2022incorporating}, and synthetic data \cite{ma2020zero, chaudhary2023exploring, dai2022promptagator} to improve the capability of models.
Recent studies \cite{rocketqa_v2, AR2} also show that joint training of the retriever and reranker can further improve their effectiveness.
In terms of encoding strategy, single-vector representation models \cite{ANCE} encode a given text as a single vector, and multi-vector representation models \cite{colbert} use multiple vectors to improve expressiveness.
Recent sparse representation models \cite{SPLADE, SPLADE++} use sparse lexical representations based on the logits of the masked language model layer of PLMs, which enables a natural query and document expansion. 

Despite their effectiveness, they require fine-tuning with massive labeled data to be adapted to the new domain corpus.
As a plug-and-play framework, \proposed complements the above approaches using a topical taxonomy without resorting to the labeled data.
% As a complementary direction for the above approaches, this work explores a systematic way to complement the existing models using topical taxonomy.

% \smallsection{Improving retrieval with auxiliary corpus knowledge}
\smallsection{Retrieval with auxiliary corpus knowledge}
These techniques aim to improve retrieval by exploiting knowledge of the target corpus.
% This direction of research is dedicated to exploiting knowledge of the target corpus.
One notable approach is pseudo-relevance feedback (PRF), which utilizes the top-ranked results from an initial retrieval to enhance the semantic matching process.
The existing methods have exploited key terms \cite{Multi-PRF, kim2022collective}, text segments \cite{BERT-QE}, and documents \cite{Dense-PRF} from top-ranked results as an additional context to complement the query.
Recently, \cite{mao2021generation, mackie2023generative} have directly utilized knowledge stored in the PLMs for query expansion.
Despite their effectiveness in filling missing contexts, their effectiveness is often limited in theme-specific applications due to the suboptimal initial retrieval quality and the need for domain-specific knowledge.

Another approach leverages inter-document similarity via a \textit{corpus graph} whose nodes are documents and edges connect most similar documents. 
Under the assumption that similar documents tend to be relevant to the same query, \cite{GAR} adapts the candidate document set for reranker using the nearest neighbors in the graph.
\cite{LADR} first uses lexical retrieval to obtain seed documents and uses the graph to gradually expand the search space for retrievers.
Lastly, there have been a few attempts to use topic information for retrieval. 
\cite{topic_IR1, topic_IR2} combine LDA \cite{LDA} with statistic-based retrieval, and \cite{TASB, TASB-study} use topics for a balanced batch construction.
\cite{li2021topic} incorporates topic information into a word embedding-based model.
However, there has been no attempt to exploit the high-quality corpus taxonomy.

% Most methods leveraging topic information \cite{topic_IR1, topic_IR2, li2021topic} are tailored for traditional retrieval models (e.g., statistic-based) and are difficult to apply to PLM-based~models.

It is worth noting that there exist a few attempts to use external knowledge (e.g., knowledge base) \cite{dong2022incorporating}.
We focus on exploiting knowledge of the target corpus, and \proposed can be combined with the external knowledge-based models as well. 
% We provide related work for topic mining and taxonomy completion in Appendix~\ref{A:relatedwork}.

\section{METHODOLOGY}
\label{sec:method}

We present \textbf{To}pical \textbf{T}axonomy-\textbf{E}nhanced \textbf{R}etrieval (\proposed) framework.
We first explain how \proposed bridges the given taxonomy with the target corpus in the training phase (\cref{subsec:TTER_training}), then present how \proposed enhances PLM-based retrieval in the inference phase (\cref{subsec:method_inference}).
% \proposed does not require labeled data from the target corpus, but it can also effectively leverage the labeled data (Appendix \ref{subsec:label}).
The overview of \proposed is presented in Figure \ref{fig:method}.

\subsection{Topic Class Relevance Learning}
\label{subsec:TTER_training}
The taxonomy reveals the latent structure of the whole corpus.
To exploit it for retrieval, we first connect the corpus-level knowledge to individual documents.
We formulate this step as an unsupervised \textit{multi-label classification}, assessing the relevance of each document to each topic class without document-topic labels.
% As it is infeasible to obtain topic annotation for all documents, we assume that no ground-truth topic labels are given.
We first introduce our silver label generation strategy (\cref{subsec:method_silver}).
Then, we propose a new training strategy to produce rich and reliable training signals (\cref{subsec:method_learning}).
Lastly, we summarize the overall training process (\cref{subsec:algo}).

\subsubsection{\textbf{Taxonomy-guided silver label generation}}
\label{subsec:method_silver}
As the first step, we seek to identify a small candidate set of relevant classes for each document, which will be our silver labels for training.
Utilizing the hierarchical structure of topic classes, we introduce a top-down approach that \textit{recursively assigns} documents to the child node with the highest similarity, gradually narrowing down the topic.
Given a document, we start from the ``root node'' and compute its similarity to each child node. 
The document is then assigned to the child node with the highest similarity.\footnote{
Although a document can cover multiple topics, we assign it to the most similar child node to generate reliable silver labels. Indeed, we observe increasing the assignment number easily leads to degraded training efficacy.}
% 
% In Sec. \ref{subsec:method_learning}, we introduce a new strategy to explore unlabeled but relevant topic classes during the training.} 
This assignment process recurs until it reaches leaf nodes.
Once every document has been assigned, we apply a \textit{filtering step} to retain only reliable labels for each document.

\smallsection{Document-class similarity computation}
For each document $d$, we compute its similarity with a child node $c_j$ by considering all phrases related to $c_j$. 
Let $P^{\mathcal{T}}_{j}$ denote the union of all phrases from the subtree having $c_j$ as a root node.
The similarity is computed by considering both lexical aspect ($\operatorname{sim}_{L}$) based on statistics and semantic aspect ($\operatorname{sim}_{S}$) based on PLMs, which are defined as:
% Following the previous studies of topic discovery \cite{zhang2023effective, lee2022taxocom, zhang2022seed}, the relevance is computed considering both statistics and PLM-based representations.
\begin{equation}
\vspace{-0.05cm}
\begin{aligned}
    {sim}_{L}(d, c_j) &= \frac{1}{|P^{\mathcal{T}}_{j}|}\sum_{p\in P^{\mathcal{T}}_{j}} \operatorname{tf}(p,d) \cdot \operatorname{idf}(p)\\
    {sim}_{S}(d, c_j) &= \frac{1}{|P^{\mathcal{T}}_{j}|}\sum_{p\in P^{\mathcal{T}}_{j}} \operatorname{cos}(\mathbf{h}_p, \mathbf{h}_d),
\end{aligned}
\vspace{-0.05cm}
\end{equation}
where $\mathbf{h}_p$ and $\mathbf{h}_d$ denote representations from PLM for a phrase $p$ and a document $d$.
%\footnote{We employ BERT \cite{BERT} and apply mean pooling to obtain the final representation.}
The lexical and semantic similarity can reveal complementary aspects;
lexical matching has strengths in handling domain-specific terminologies that rarely exist in the general corpus, and semantic matching excels in capturing broader contextual meanings while flexibly handling non-exact matching terms.

To jointly consider the two aspects, we adopt the ensemble score based on the reciprocal rank \cite{zhang2023effective, zhang2022seed}.
%\footnote{This corresponds to the geometric mean of the two reciprocal ranks ($\rho \rightarrow 0$) \cite{zhang2022seed}.}
By ranking all child nodes in descending order of ${sim}_{L}$ and ${sim}_{S}$, each node will have two rank positions ${rank}_{L}$ and ${rank}_{S}$, respectively. 
The overall similarity is:
\begin{equation}
\begin{aligned}
    {sim}_{O}(d, c_j) = \left(\frac{1}{2}\left(\frac{1}{{rank}_{L}(c_j)}\right)^{\rho} + \frac{1}{2}\left(\frac{1}{{rank}_{S}(c_j)}\right)^{\rho} \right)^{1/{\rho}},
\end{aligned}
\end{equation}
where $0 < \rho \leq 1$ is a constant. We set $\rho=0.1$.
In this ensemble score, a high position in one ranking can largely compensate for a lower position in another, thus allowing for the joint consideration of two distinct aspects. 

\smallsection{Filtering step}
After the class assignment for all documents, we apply a filtering step to only retain assignments with high similarity.
For each class node, we keep documents whose similarity exceeds the median similarity of all documents assigned to the class. 
If a document is filtered out from a certain node, the document is also removed from all its child nodes, ensuring hierarchical consistency.\footnote{We apply the filtering to nodes at the second or deeper levels to ensure every document has at least one class assignment (the root node has level 0).}
Finally, for each document $d$, we obtain silver labels $\mathbf{y}^{s}_{d} \in \{0,1\}^{|\mathcal{C}|}$,
where $y^{s}_{dj}=1$ if $c_j$ is assigned to $d$, otherwise $0$.

% (denoted as $D(c_j)$): ${rel(d, c_j)} \geq {median}({rel}(d', c_j) \mid d' \in D(c_j))$.
% By doing so, we can retain reliable class information and also filter out overly specific topic classes for each document.

\subsubsection{\textbf{Class relevance learning}}
\label{subsec:method_learning}
Based on the obtained silver labels, we train a \textit{class relevance estimator} that predicts relevance between a document and topic classes. 
For effective training, we propose a new \textit{collective topic knowledge distillation} strategy designed to complement incomplete silver labels.

\smallsection{Class relevance estimator}
As developing a novel architecture is not the focus of this paper, we employ the existing methods to encode documents and class nodes.
For the \textit{text encoder}, we use BERT \cite{BERT} with mean pooling to obtain $\mathbf{h}_d$ for each document $d$.
For the \textit{topic class encoder}, we adopt graph convolutional networks (GCNs) \cite{GCN} to incorporate both semantic and structural information.
For each class node $c_j$, we first obtain its ego graph that includes its $L$-hop neighboring nodes and apply GCN layers to propagate node features over the taxonomy structure.
Each node feature is initialized by the BERT representation of its class name.
After stacking $L$ GCN layers, we use the representation of the ego node, denoted as $\mathbf{c}_j$, as the final class representation.

Then, we calculate the topic class-document relevance by bilinear interaction between their representations, i.e., $\hat{y}_{dj}=\sigma(\mathbf{c}^\top_j\mathbf{M}\mathbf{h}_d)$, where $\mathbf{M}$ is a trainable interaction matrix and $\sigma(\cdot)$ is the sigmoid function.
The estimator is trained by the binary cross-entropy loss:
\begin{equation}
\begin{aligned}
\label{eq:bce}
    \min_{\mathbf{M},\, \theta_{GCN}} \mathcal{L} = - \sum_{d \in \mathcal{D}}\sum_{c_j \in \mathcal{C}} y_{dj} \log \hat{y}_{dj} + (1 - y_{dj}) \log (1 - \hat{y}_{dj}).
\end{aligned}
\end{equation}
In the early stages of training, we use the silver labels $\mathbf{y}_{d} = \mathbf{y}^{s}_{d}$.
As training progresses, we exploit collective topic labels $\mathbf{y}_{d} = \mathbf{y}^{c}_{d}$ obtained from similar documents, which we will introduce below.

\smallsection{Collective topic knowledge distillation (CKD)}
As the silver labels incompletely reveal true relevance classes, relying solely on them leads to suboptimal estimation.
As a solution, we propose CKD, designed to complement the incomplete labels.
Our core idea is that the topic distribution of a document can be inferred from semantically similar documents.
Specifically, (1) Using each document $d$ as a query, we \textit{retrieve} a small subset of semantically similar documents $\mathcal{D}_{d}$ from the corpus.
To accurately retrieve $\mathcal{D}_{d}$, we consider both semantic similarity (via a dual-encoder) and topical relatedness (via our estimator). This will be explained in Sec \ref{method_TRM}.
(2) We compute the class relevance distributions for each retrieved document, i.e., $\{\mathbf{\hat{y}}_{d'} \mid d' \in \mathcal{D}_{d}\}$.
(3) By averaging the predicted distributions, we generate collective relevance labels $\mathbf{y}^{c}_{d} \in (0,1)^{|C|}$.

Unlike $\mathbf{y}^{s}_{d}$ which consists of binary values, $\mathbf{y}^{c}_{d}$ reveals the soft probability of a document's relevance to each class. 
Notably, $\mathbf{y}^{c}_{d}$ reveals topic classes that are highly pertinent to documents similar to $d$, providing rich supervision not included from $\mathbf{y}^{s}_{d}$.
Moreover, this collective knowledge distills more stable and reliable signals than using individual predictions for pseudo-labeling, as done in conventional self-training \cite{self-training, pseudo-labeling}.
It is worth noting that $\mathbf{y}^{c}_{d}$ gets refined during the training.
That is, the topic estimator is improved with collective knowledge, which again results in more accurate discovery of similar documents and their topic~distributions.

\begin{algorithm}[t]
\small
\SetKwInOut{Input}{Input}
\SetKwInOut{Output}{Output}
\Input{A target corpus $\mathcal{D}$, a corpus topical taxonomy $\mathcal{T}$, \\a retriever $f$, a update period $t$}
\Output{Trained class relevance estimator $g$}
Randomly initialize the training parameters of $g$ \\
Generate silver labels $\mathbf{y}^s_d$ for all $d \in \mathcal{D}$ \COMMENT{\textit{\cref{subsec:method_silver}}}\\
Warm-up the estimator $g$ using only $\mathbf{y}^s_d$ \COMMENT{\textit{Eq.\ref{eq:bce}}}\\
\BlankLine
\tcc{Collective knowledge distillation}
\For{$e=1,... ,\text{ }epoch_{max}$  }{
\ForEach{$d \in \mathcal{D}$}{
\If{$e\,\%\, t == 0$}{
    % \tcc{Collective label generation}
    Retrieve $\mathcal{D}_d$ using $f$ with CRM \COMMENT{\textit{\cref{method_TRM}}}\\
    Compute class relevance as $\{\mathbf{\hat{y}}_{d'} \mid d' \in \mathcal{D}_{d}\}$ using $g$\\
    Obtain collective labels $\mathbf{y}^c_d = \operatorname{AVG}(\{\mathbf{\hat{y}}_{d'} \mid d' \in \mathcal{D}_{d}\})$\\
}
Train the estimator $g$ using $\mathbf{y}^c_d$ \COMMENT{\textit{Eq.\ref{eq:bce}}}\\
}
}
\caption{Training algorithm of \proposed.}
\label{algo:algo}
\end{algorithm}

\subsubsection{\textbf{Training algorithm of \proposed}}
\label{subsec:algo}
The training process of \proposed is provided in Algorithm \ref{algo:algo}.
% We first initialize the class relevance estimator $g$, which has the training parameters for the class encoder (i.e., $\theta_{GNN}$) and bilinear interaction matrix (i.e., $\mathbf{M}$).
We generate the silver labels $\mathbf{y}^s_d$ and use them to warm up $g$.
After the warm-up, we train $g$ with the collective knowledge distillation.
In specific, for each document $d$, we retrieve a set of similar documents $\mathcal{D}_d$ with CRM, and generate the collective labels $\mathbf{y}^c_d$ by averaging their class relevance distributions.
In practice, we update the collective topic labels $\mathbf{y}^c_d$ every $t$ epochs, which makes the training process more efficient and robust. 
In this work, we set $|\mathcal{D}_d|=10$ and $t=25$.

\subsection{Topical Taxonomy-Enhanced Retrieval}
\label{subsec:method_inference}
We present how \proposed improves PLM-based retrieval at the inference phase.
\proposed consists of three strategies to complement the existing retrieve-then-rerank pipeline.
Each strategy is designed to gradually focus on fine-grained ranking, as shown in Figure \ref{fig:method}.

% \vspace{-0.05cm}
\smallsection{Class relevance estimation}
After training, for every document $d$ in the corpus, we compute its topic class relevance as~$\hat{\mathbf{y}}_d$.
Considering each document only covers a small subset of topics within the corpus, we focus on classes with high relevance.
To indicate these \textit{relevant classes}, we introduce a binary indicator vector $\hat{\mathbf{b}}_d \in \{0,1\}^{|\mathcal{C}|}$, where $\hat{b}_{dj}=1$ denotes that $d$ has a certain degree of relevance to $c_j$, otherwise $\hat{b}_{dj}=0$.
We recursively retain the top $m\%$ classes for each level of taxonomy by setting the corresponding elements of $\hat{\mathbf{b}}_d$ as $1$.
If a class is not retained at a higher level, all its child classes are not retained as well, ensuring hierarchical consistency.
We set $m=10$.
% We recursively retain classes whose scores exceed the $m$-quantile for each level of taxonomy by setting the corresponding elements of $\hat{\mathbf{b}}_d$ as $1$.
% If a class is not retained at a higher level, all its child classes are not retained as well, ensuring hierarchical consistency.
% We set $m=0.9$.
% We recursively retain classes with top $m\%$ scores for each level of taxonomy by setting the corresponding elements of $\hat{\mathbf{b}}_d$ as $1$.
% If a class is not retained at a higher level, all its child classes are not retained as well, ensuring hierarchical consistency.
% We set $m=10$.
At test time, for a given query $q$, we obtain $\hat{\mathbf{y}}_q$ and $\hat{\mathbf{b}}_q$ in the same~way.

% Starting from the root node, we recursively retain the top $m\%$ relevant child nodes by setting the corresponding elements of $\hat{\mathbf{b}}_d$ as $1$.

\subsubsection{\textbf{Search space adjustment (SSA) to reduce search space.}}
\label{method_SSA}
The topic class relevance reveals the central subjects of each document, providing a snapshot of its main focus.
Before applying the PLM-based retrievers, we seek to \textit{filter out} a large number of irrelevant documents having little topic class overlap with the query. 
This step can benefit subsequent retrieval by reducing the search space while preserving topically relevant documents that may otherwise be overlooked by PLM-based retrievers.
For search space reduction, lexical models (e.g., BM25) are mostly considered due to their high efficiency \cite{LADR, thakur2021beir}.
We expect topic-based SSA can have strengths in identifying relevant documents, compared to using lexical similarity based on word overlap.

As topics are discrete categories, we can efficiently compute the topic overlap using the binary vectors.
In specific, we compute the degree of topic overlap between the query and each document using \textit{bitwise operations}: $\operatorname{Popcount}(\operatorname{AND}(\hat{\mathbf{b}}_q, \hat{\mathbf{b}}_d))$.
Then, we filter out documents with low degrees of overlap, obtaining the reduced search space $\mathcal{D}^{SSA}_q$.
The size of this space can be determined empirically.
We continue subsequent retrieval on $\mathcal{D}^{SSA}_q$ instead~of~$\mathcal{D}$.

\textbf{Remarks.} Compared to using real-valued vectors $\hat{\mathbf{y}}_*$, the proposed SSA is more efficient as it uses bitwise operations of binary vectors, largely reducing the need for floating-point operations.
It can be further accelerated using multi-index hashing for binary codes \cite{binary_indexing}.
As search speed acceleration is a distinct research topic, we focus on the accuracy aspect in this work.
% Our topic-based SSA can be further combined with lexical models as a kind of hybrid method. We leave further exploration for future study.

% whose fine-grained rankings are decided by a subsequent reranker.
\subsubsection{\textbf{Class relevance matching (CRM) for retriever}}
\label{method_TRM}
The first-stage retriever aims to find a set of candidate documents~$\mathcal{D}_q$. 
In this step, we exploit \textit{topical relatedness} of the query and document, which is the similarity between distributions of the relevant topic classes.
Topical relatedness focuses on the relevance of the central subjects of the input texts identified using the class estimator.
This can help to handle lexical mismatches and fill in missing contexts, providing a complementary aspect to semantic similarity. % by the dual-encoder.
% This other dimension of relevance, derived from topic distribution through theme-specific taxonomy, provides a complementary aspect to semantic similarity by the dual-encoder.
Formally, we retrieve $\mathcal{D}_q$ based on $s(q,d)$, considering both semantic similarity from dual-encoder $s_{de}(q,d)$ and topical relatedness $s_{CRM}(q,d)$:
\begin{equation}
\begin{aligned}
s(q,d) = combine(s_{de}(q,d), \,s_{CRM}(q,d)),\\
s_{CRM}(q,d) = sim(\hat{\mathbf{y}}_q \odot \hat{\mathbf{b}}_q, \,\hat{\mathbf{y}}_d \odot \hat{\mathbf{b}}_d).
\end{aligned}
\end{equation}
We obtain the relevant class distribution using element-wise multiplication, denoted as $\odot$, with the binary vector.
Then, we compute the similarity between the query and document distributions using the $sim(\cdot, \cdot)$ function, where we use inner-product.
% In practice, we can only save and check scores of the relevant classes (i.e., $\{\hat{y}_{*j}|\hat{b}_{*j}=1\}$) to improve efficiency.
$combine(\cdot, \cdot)$ denotes a function to consolidate two scores, and we adopt a simple addition with rescaling via z-score normalization.\footnote{A hyperparameter can be also used to balance two scores of varying magnitudes. 
% However, we consistently achieved satisfactory results using simple normalization.
}
We also explored reflecting the granularity of topics by exclusively focusing on broad or specific topic classes in CRM.
However, considering the overall topic distribution, encompassing both broad and specific topics, proved most effective.
Please refer to \cref{A:subsec_sup} for a detailed~study.

% However, considering the overall topic distributions including both broad and specific ones proved most effective. 
% However, considering the overall topics including both broad and specific topics of queries and documents proved most effective. 

\subsubsection{\textbf{Query enrichment by core phrases (QEP) for reranker.}}
\label{subsubsec:method_qep}
In this last stage, a reranker reorders $\mathcal{D}_q$ based on their fine-grained relevance to the query $q$.
Since $\mathcal{D}_q$ already have similar relevant classes via CRM, in this step, we delve deeper into each topic by focusing on class-related phrases.
As discussed earlier, users familiar with a domain often omit contexts in their queries, which makes it difficult to find accurate relevance.
To address this, we use phrase-level knowledge to enrich queries.

A remaining question is how to identify phrases to complement a given query.
QEP is built upon the relevance model philosophy \cite{lavrenko2017relevance}, which assumes that both a query and its relevant documents are generated from a shared underlying relevance model.
Although the true relevance model behind the query is unknown, it can be inferred from the most relevant documents obtained from retrieval \cite{Dense-PRF, Multi-PRF, BERT-QE}. 
Based on this idea, we identify \textit{core phrases} to enrich the query using both the topic class knowledge and top-ranked retrieved documents.
From the set of relevant class phrases $\{P_j | \hat{b}_{qj}=1\}$, we collect the top-$k$ core phrases $P^q$ that most frequently appear in top-ranked documents.\footnote{The phrases related to each class ($P_j$) are provided in the topical taxonomy (\cref{subsec:concept}). The frequencies of phrases for each document are pre-computed.}
Then, we enrich the original query with $P^q$ and use the enriched query for inference:
\begin{equation}
\label{eq:tmplt}
    s_{ce}(q,d) = f_{\phi}(q^{QEP}, d), \quad
    q^{QEP} = [q; \texttt{TMPLT}(P^q)],
\end{equation}
% \vspace{-0.000001pt}
where $\texttt{TMPLT}$ denotes the template of the hard prompt. 
In this work, we use $\texttt{TMPLT}(P^q) = ``\textit{, relevant topics: }\{P^q\}$''.
We also tried using just the topic class names (i.e., the most salient phrase) for query enrichment. 
However, we found that they are often too coarse-grained, thus bringing limited information for fine-grained rankings.

% \subsection{\proposed with Labeled Data}
% \label{subsec:label}
% While \proposed mainly focuses on scenarios without available labeled data from the target corpus, it can also benefit from harnessing $(q,d)$ labels. 
% First, \proposed can directly use $q$ for class relevance learning (Eq.\ref{eq:bce}) by treating it as an additional document.
% By using the class labels of its relevant document $d$, we can reflect their relevance into the class estimator. 
% Also, the enriched queries by \proposed can be directly used to enhance the fine-tuning of the retriever and reranker.
% That is, we use $q^{QEP}$ instead of $q$ as model input, where the fine-tuning follows the standard contrastive learning \cite{DPR}.

\section{Experiments}
\label{sec:experimentsetup}
We experiment to answer the following research questions: 
\begin{researchquestions}%\vspace{-\topsep}
\item How does each strategy of \proposed affect retrieval accuracy? 
\item How does \proposed compare with other techniques that use auxiliary corpus knowledge in terms of retrieval accuracy?
\item Is \proposed compatible with a variety of PLM-based retrieval models used for each retrieval stage?
\item How does each proposed component affect \proposed?
\item How does the labeled data affect the effectiveness of \proposed?
\item How sensitive is \proposed to the taxonomy quality?
\end{researchquestions}%\vspace{-\topsep}

% \shivam{How does \proposed compare with other techniques that use auxiliary corpus knowledge in terms of retrieval accuracy?}
% What is the impact of \proposed on retrieval accuracy compared to other techniques that use auxiliary corpus knowledge?

\subsection{Experimental Setup}
% \noindent
% We provide details of setup and implementation in Appendix \ref{A:exp_detail}.

\subsubsection{\textbf{Dataset}}
We simulate two theme-specific applications: (1) academic paper search using SCIDOCS dataset \cite{thakur2021beir, SPECTER}, (2) product search in e-commerce using Amazon ESCI dataset \cite{ESCI}.\footnote{\url{https://github.com/amazon-science/esci-data}}
SCIDOCS dataset is widely used as a benchmark dataset evaluating the zero-shot prediction capability of retrieval models \cite{thakur2021beir}.
Amazon ESCI dataset is adopted from KDD Cup 2022-Task 1.
We use the English (US) data and treat ‘E (exact match)’ as the relevant relation.
Each document contains the product title, product description, and product bullet points.
We focus on scenarios with no training labels from the target corpus in \cref{sec:results} and analyze the impacts of labeled data in \cref{subsec:label_results}.
Table \ref{tab:statistics} summarizes the data statistics.

\begin{table}[h]
\caption{Data statistics of two datasets. Avg. D/Q indicates the average number of relevant documents per query.}
\renewcommand{\arraystretch}{0.7}
\label{tab:statistics}
\resizebox{0.8\linewidth}{!}{
\begin{tabular}{ccc}\toprule
& \multicolumn{1}{c}{Academic domain} & \multicolumn{1}{c}{Product   domain} \\ \midrule\midrule
\#Corpus & 25,657 & 601,354 \\
\#Training query & - & 20,888 \\
\#Test query & 1,000& 8,956 \\
Avg. D/Q & 4.9 & 8.83 \\ \bottomrule
\end{tabular}}
\vspace{-0.3cm}
\end{table}

\subsubsection{\textbf{Corpus Topical taxonomy construction}}
\label{A:taxoconstruction}
The corpus topical taxonomy is obtained by applying the existing taxonomy completion technique \cite{lee2022taxocom} on a seed taxonomy.
For the seed taxonomy, we utilize the fields of study hierarchy from Microsoft Academic Graph \cite{MAG_FS} (for the academic domain) and Amazon store taxonomy\footnote{\url{https://www.amazonlistingservice.com/blog/amazon-store-taxonomy-organization}} sourced from \url{Amazon.com} (for the product domain).
Each seed taxonomy mirrors user interest in each application.
The former reflects researchers' inclination towards structuring academic concepts and studies, while the latter embodies customers' interest in browsing and selecting products. 
Table \ref{tab:taxo_statistics} provides the statistics of the constructed taxonomies.

\begin{table}[h]
\caption{Taxonomy statistics of two datasets. }
\renewcommand{\arraystretch}{0.7}
\label{tab:taxo_statistics}
\resizebox{0.75\linewidth}{!}{
\begin{tabular}{ccc}\toprule
& \multicolumn{1}{c}{Academic domain} & \multicolumn{1}{c}{Product domain} \\ \midrule\midrule
\#Topic classes &  4,028 &  14,954\\
\#Edges & 8,445 & 18,360\\
Depth & 5 & 10 \\\bottomrule
\end{tabular}}
\vspace{-0.4cm}
\end{table}

\subsubsection{\textbf{Retrieval setup and metrics}}
Without \proposed, the retrieval process follows the standard (1) retrieval, (2) reranking pipeline (\cref{subsec:concept}).
\proposed has an added SSA step to reduce the initial search space. 
Thus, with \proposed, the retrieval process follows (1) SSA, (2) retrieval (with CRM), and (3) reranking (with QEP).

For each retrieval stage, we employ the metrics and retrieval sizes widely used in the prior literature \cite{GAR, LADR}. 
In the first-stage retrieval, which aims to retrieve all potentially relevant documents, recall is mostly considered, with retrieval sizes typically ranging from $100$ to $1000$. 
For the second-stage reranking, more sophisticated metrics (e.g., NDCG, MAP) that consider the detailed rankings of relevant documents are widely employed, with smaller retrieval sizes within $100$. 
For SSA and the first-stage retrieval, we use Recall (R@$K$).
For the second-stage reranking, we additionally use NDCG (N@$K$) and MAP (MAP@$K$).
In our experiments, we set the size of search space adjustment $|\mathcal{D}^{SSA}_q|=2500$ and the candidate set for reranking $|\mathcal{D}_q|=100$ \cite{GAR}.

\begin{figure*}[t]
\centering
\includegraphics[width=0.247\textwidth]{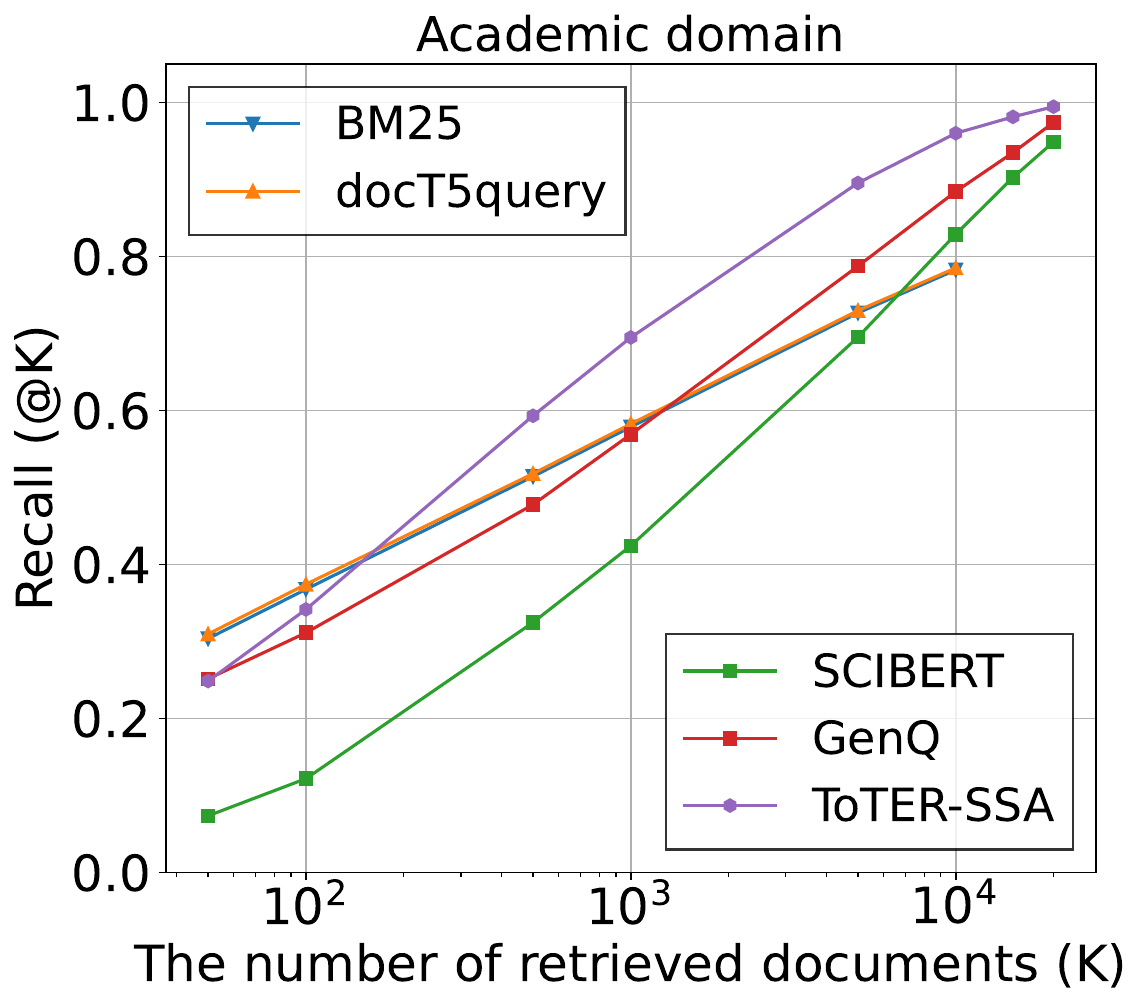}
\includegraphics[width=0.247\textwidth]{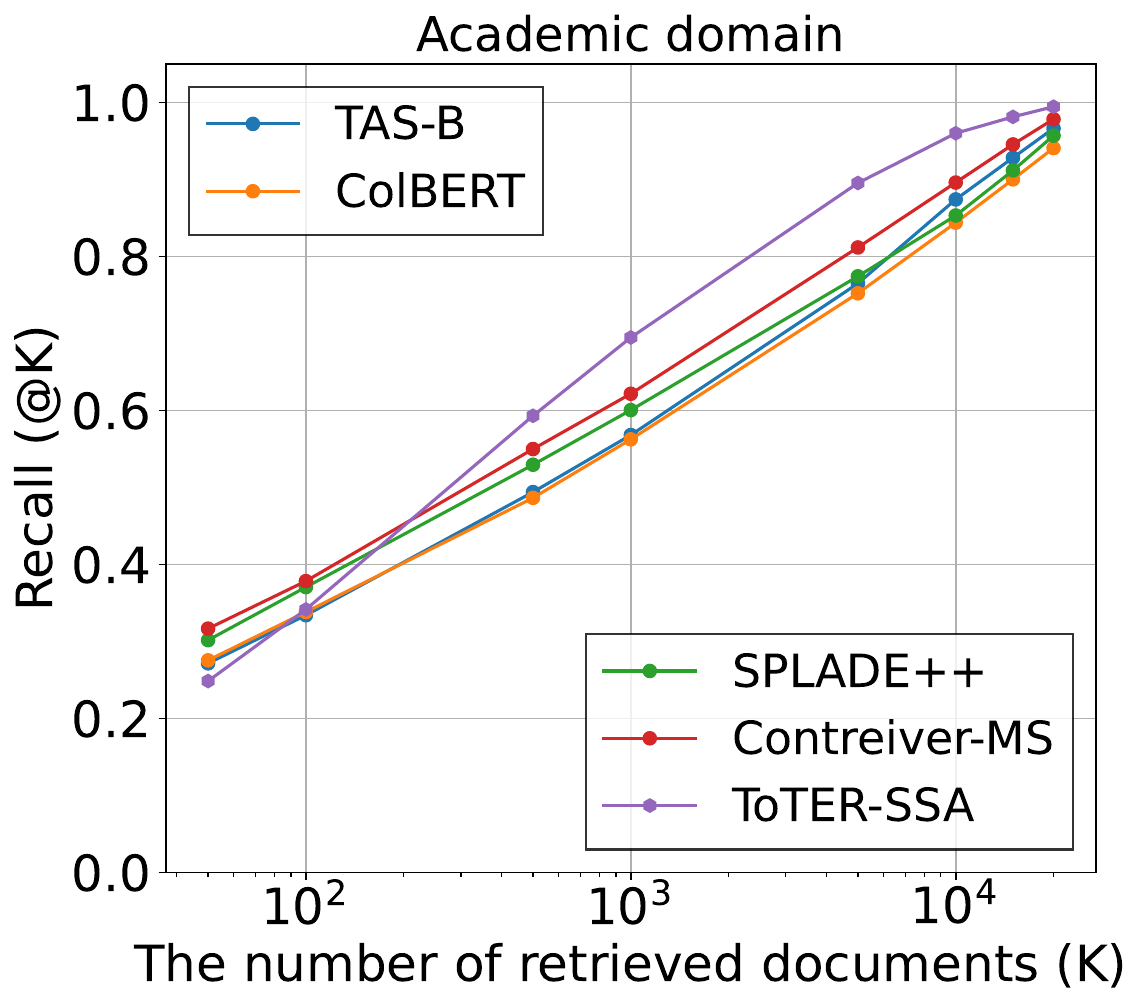}
\includegraphics[width=0.247\textwidth]{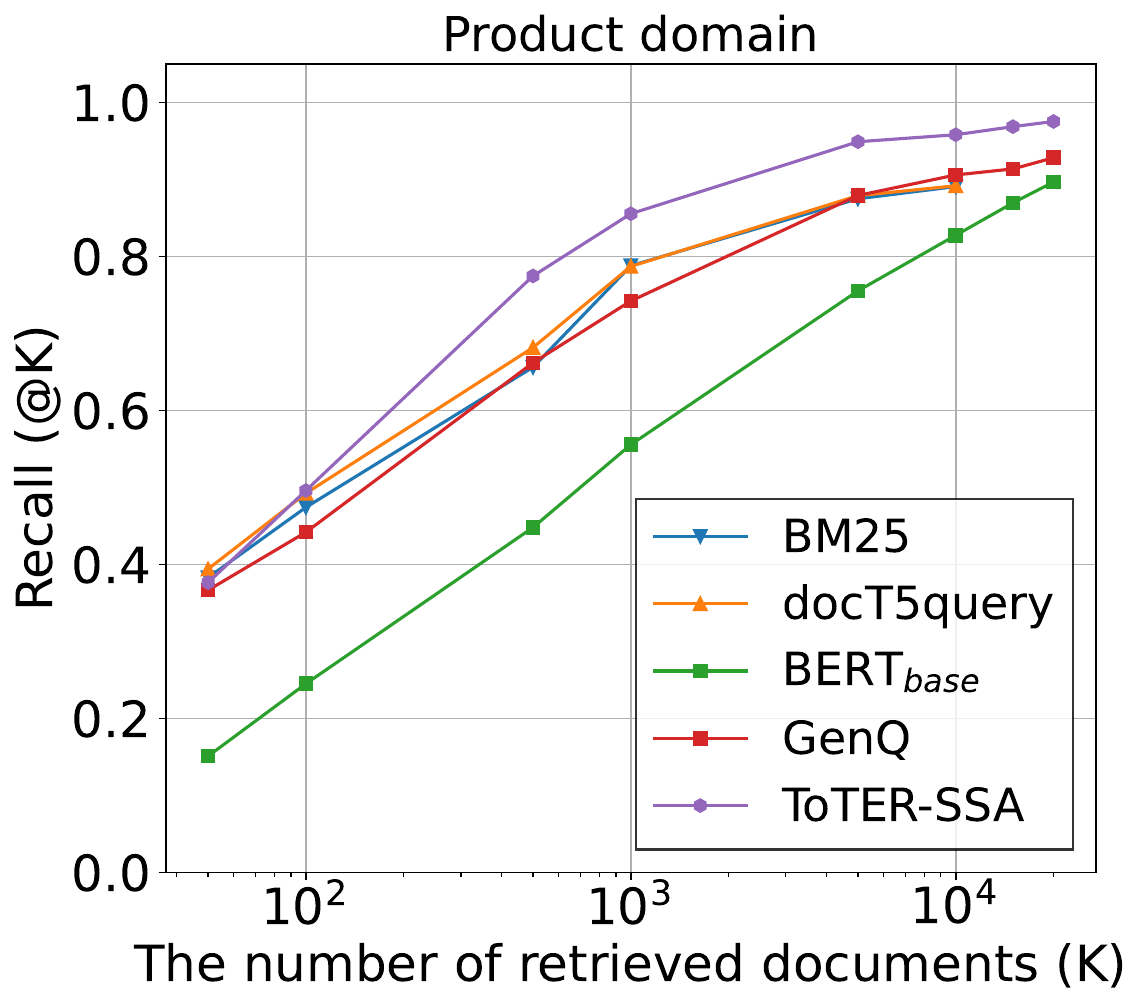}
\includegraphics[width=0.247\textwidth]{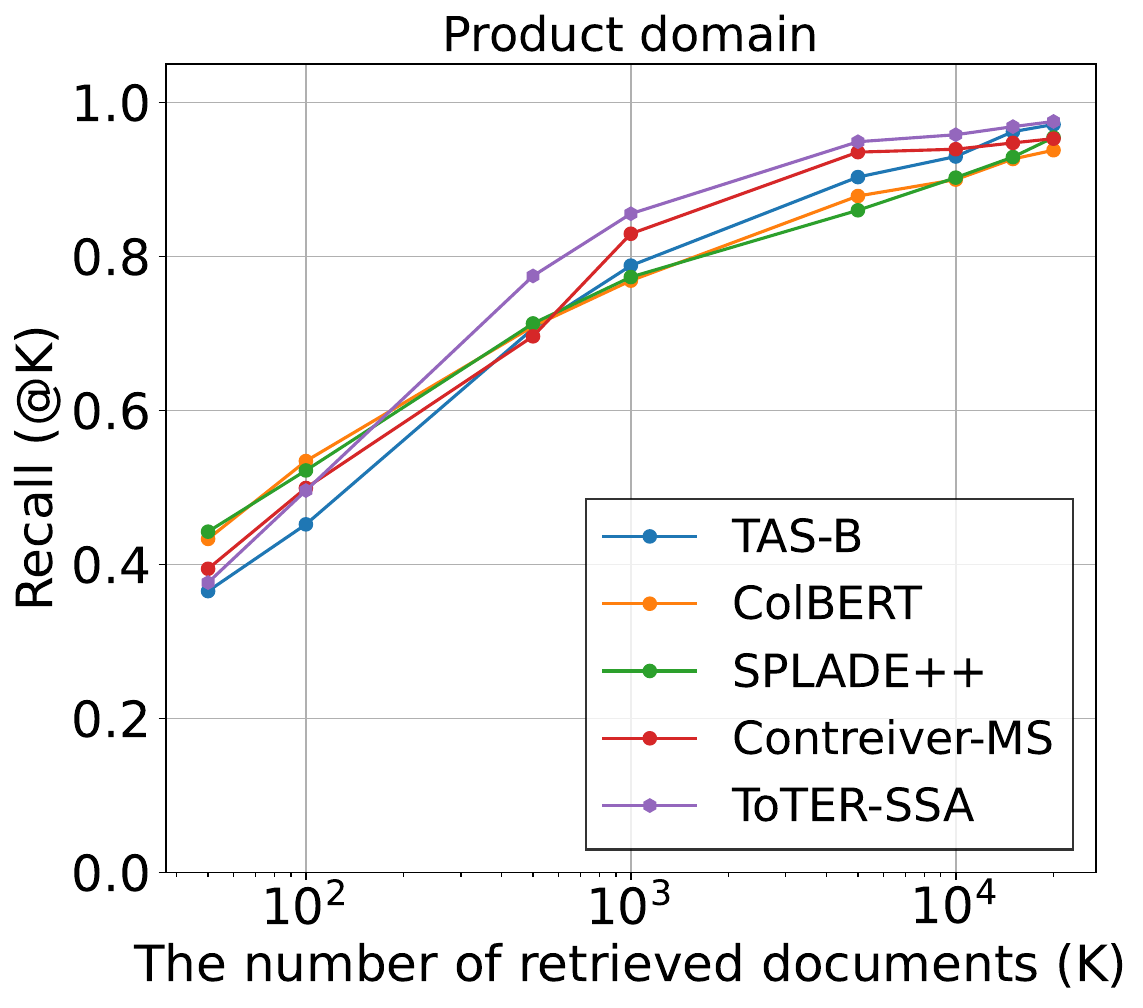}
\caption{Search space adjustment performance comparison on the academic and product domain.}
\label{fig:SSA}
% \vspace{-0.4cm}
\end{figure*}

\subsection{Results and Analysis (RQ1,RQ2, and RQ3)}
\label{sec:results}
To validate the effectiveness of each strategy of \proposed, we provide comparisons with various baselines for each retrieval stage.
% We provide a detailed analysis for each stage of retrieval.

% \vspace{-0.18cm}
\subsubsection{\textbf{Initial search space adjustment}}
We first assess the efficacy of SSA in filtering out irrelevant documents from the corpus.
As this can be seen as a kind of retrieval with a large retrieval size, we compare various retrieval methods.
Note that SSA does not require backbone retrieval models, as it only uses the~class~estimator.

\smallsection{Compared methods} 
(a) Lexical and sparse model: BM25 \cite{BM25}, docT5query \cite{DocT5query}, 
(b) pre-trained PLMs for each corpus: SCIBERT \cite{SCIBERT} \& BERT-base \cite{BERT},
(c) unsupervised domain adaptation: GenQ \cite{thakur2021beir}, 
(d) PLM-based retriever: \CTR-MS \cite{CTR}, TAS-B \cite{TASB}, ColBERT \cite{colbert}, SPLADE++ \cite{SPLADE++}.
GenQ uses synthetic data generated for the target corpus\footnote{We fine-tune \CTR-MS using BM25 negatives on the synthetic data.}, and other retrievers in (d) are fine-tuned using massive labeled data from the general domain \cite{nguyen2016ms}.

% \usepackage{booktabs}
% \usepackage{multirow}
\begin{table*}[t]
\caption{Retrieval performance comparison on the academic and product domain. Significant differences with the baseline (i.e., retrieval without using \proposed framework) are marked with * (p-value < 0.05 in the one-sample t-test).}
\small
\renewcommand{\arraystretch}{0.75}
\resizebox{1.\linewidth}{!}{
\begin{tabular}{@{}cclccccccccc@{}}
\toprule
 & \multirow{2}{*}{Search space} & \multirow{2}{*}{Method} & \multicolumn{3}{c}{Contriever-MS} & \multicolumn{3}{c}{SPLADE++} & \multicolumn{3}{c}{ColBERT} \\ \cmidrule(lr){4-6} \cmidrule(lr){7-9} \cmidrule(lr){10-12}
 &  &  & R@100 & R@500 & R@1000 & R@100 & R@500 & R@1000 & R@100 & R@500 & R@1000 \\ \midrule\midrule
\multirow{12}{*}{\makecell[c]{Academic\\domain}} &  & Retriever & 0.3783 & 0.5498 & 0.6216 & 0.3705 & 0.5294 & 0.6004 & 0.3382 & 0.4864 & 0.5624 \\
 &  \multirow{4}{*}{\begin{tabular}[c]{c}Entire corpus\\ $(\mathcal{D})$\end{tabular}} & w/ \QE & 0.3846 & 0.5543 & 0.6280 & \textbf{0.3911} & 0.5523 & 0.6193 & 0.3484 & 0.4991 & 0.5783 \\
 &  & w/ PRF & 0.3815 & 0.5510 & 0.6266 & 0.3852 & 0.5441 & 0.6146 & 0.3484 & 0.5011 & 0.5860 \\ 
 &  & w/ GAR & \textbf{0.3848} & 0.5499 & 0.6218 & 0.3772 & 0.5336 & 0.6033 & 0.3486 & 0.4912 & 0.5629 \\
 &  & w/ LADR & 0.3806 & \textbf{0.5626} & \textbf{0.6302} & 0.3763 & \textbf{0.5577} & \textbf{0.6264} & \textbf{0.3501} & \textbf{0.5339} & \textbf{0.6115} \\ 
  &  & w/ \TQE & 0.3835&	0.5491&	0.6261&	0.3672	&0.5271&	0.5995&	0.3393&	0.4845&	0.5581 \\ \cmidrule(lr){2-12}
 &  & Retriever & 0.3887 & 0.5866 & 0.6793 & 0.3913 & 0.5809 & 0.6725 & 0.3586 & 0.5486 & 0.6502 \\
 &  \multirow{4}{*}{\begin{tabular}[c]{c}Reduced corpus\\ $(\mathcal{D}^{SSA})$\end{tabular}} & w/ \QE & 0.3940 & 0.5861 & 0.6836 & 0.4086 & 0.5942 & 0.6856 & 0.3658 & 0.5557 & 0.6553 \\
 &  & w/ PRF & 0.3915 & 0.5862 & 0.6806 & 0.4032 & 0.5896 & 0.6806 & 0.3615 & 0.5547 & 0.6593 \\
 &  & w/ GAR & 0.3928 & 0.5862 & 0.6802 & 0.4099 & 0.5966 & 0.6815 & 0.3964 & 0.5887 & 0.6810 \\
 &  & w/ \TQE & 0.3933&	0.5873&	0.6802&	0.3902	&0.5772	&0.6719&	0.3584&	0.5420&	0.6382  \\
 & &\cellcolor{gray!15}w/ \proposed-CRM &\cellcolor{gray!15} \textbf{0.4432*} &\cellcolor{gray!15} \textbf{0.6399*} &\cellcolor{gray!15} \textbf{0.7268*} &\cellcolor{gray!15} \textbf{0.4490*} &\cellcolor{gray!15} \textbf{0.6364*} &\cellcolor{gray!15} \textbf{0.7252*} &\cellcolor{gray!15} \textbf{0.4326*} &\cellcolor{gray!15} \textbf{0.6346*} &\cellcolor{gray!15} \textbf{0.7232*} \\ \midrule
\multirow{12}{*}{\makecell[c]{Product\\domain}} &  & Retriever & 0.4992 & 0.6962 & 0.8294 & 0.5220 & 0.7129 & 0.7732 & 0.5342 & 0.7091 & 0.7685 \\
&  \multirow{4}{*}{\begin{tabular}[c]{c}Entire corpus\\ $(\mathcal{D})$\end{tabular}} & w/ \QE & \textbf{0.5256} & 0.7258 & 0.8481 & \textbf{0.5582} & 0.7398 & 0.8056 & \textbf{0.5603} & \textbf{0.7593} & 0.8075 \\
  &  & w/ PRF & 0.5097 & \textbf{0.7444} & \textbf{0.8572} & 0.5244 & \textbf{0.7661} & \textbf{0.8383} & 0.5350 & 0.7331 & \textbf{0.8265} \\
 &  & w/ GAR & 0.5158 & 0.7231 & 0.8409 & 0.5455 & 0.7214 & 0.8067 & 0.5515 & 0.7421 & 0.8140 \\
 &  & w/ LADR & 0.5157 & 0.7228 & 0.8498 & 0.5377 & 0.7308 & 0.8215 & 0.5435 & 0.7228 & 0.8259 \\ 
  &  & w/ \TQE & 0.5172&	0.7343&	0.8298&	0.5334&	0.7235&	0.7924&	0.5252&	0.7072&	0.7818 \\ \cmidrule(lr){2-12}
 &  & Retriever & 0.5009 & 0.7085 & 0.8555 & 0.5231 & 0.7264 & 0.8229 & 0.5303 & 0.7401 & 0.8394 \\
 &  \multirow{4}{*}{\begin{tabular}[c]{c}Reduced corpus\\ $(\mathcal{D}^{SSA})$\end{tabular}} & w/ \QE & 0.5285 & 0.7411 & 0.8632 & 0.5608 & 0.7467 & 0.8462 & 0.5638 & 0.7602 & 0.8492 \\
 &  & w/ PRF & 0.5124 & 0.7462 & 0.8663 & 0.5311 & 0.7779 & 0.8603 & 0.5355 & 0.7649 & 0.8495 \\
 &  & w/ GAR & 0.5427 & 0.7593 & 0.8564 & 0.5606 & 0.7540 & 0.8416 & 0.5569 & 0.7634 & 0.8396 \\
&  & w/ \TQE & 0.5186&	0.7381&	0.8493&	0.5384&	0.7286&	0.8351&	0.5390&	0.7459&	0.8598  \\
 &  & \cellcolor{gray!15}w/ \proposed-CRM  &\cellcolor{gray!15} \textbf{0.5515*} &\cellcolor{gray!15} \textbf{0.7899*} &\cellcolor{gray!15} \textbf{0.8692*} &\cellcolor{gray!15} \textbf{0.5717*} &\cellcolor{gray!15} \textbf{0.7856*} &\cellcolor{gray!15} \textbf{0.8648*} &\cellcolor{gray!15} \textbf{0.5661} &\cellcolor{gray!15} \textbf{0.7997*} &\cellcolor{gray!15} \textbf{0.8625*}\\ \bottomrule
\end{tabular}
}
\label{tab:retrieval}
% \vspace{-0.2cm}
\end{table*}

\smallsection{Findings} 
Figure \ref{fig:SSA} presents recalls (R@$K$) for varying sizes of retrieved documents ($K$).
SSA consistently achieves the highest recall when $K$ is large ($\geq 10^3$), which shows its efficacy in accurately filtering out irrelevant documents.
Using class relevance learning, \proposed categorizes documents based on the theme-specific taxonomy. 
This approach aids in identifying the central subject of documents, which may not be effectively captured by the lexical and semantic similarity based on word overlap and contextual meaning. 

Among the competitors, PLM-based retrievers, fine-tuned with vast labeled data, consistently show high recalls.
On the other hand, GenQ, fine-tuned with synthetic data, shows limited performance. This result also aligns with \cite{chaudhary2023exploring}.
We find that the generated queries are often trivial and fail to reflect the domain-specific knowledge, which may lead to suboptimal results. 
Lastly, the effectiveness of SSA rapidly declines when $K$ is small ($\leq 5 \times 10^2$).
This outcome is expected, given that SSA only considers the overlap degree of relevant topic classes.

\subsubsection{\textbf{Retrieval}}
We assess the effectiveness of CRM for the first-stage retrieval. 
We select three backbone models, which show competitive performance and also represent three different encoding strategies: Contreiver-MS, SPLADE++, and ColBERT.
These correspond~to single-vector, sparse, and multi-vector representation models, respectively.
We report results for both (a) without SSA (i.e., retrieval from $\mathcal{D}$) and (b) with SSA (i.e., retrieval from $\mathcal{D}^{SSA}$).

\smallsection{Compared methods}
We compare CRM with various state-of-the-art methods to improve backbone retrievers using auxiliary corpus knowledge.
The first group leverages pseudo-relevance feedback.
\begin{itemize}[leftmargin=*]%\vspace{-\topsep}
    \item \textbf{BERT-QE} \textbf{(BQE)} \cite{BERT-QE} uses core text segments (or chucks) obtained from top retrieval results to complement the original~query.
    \item \textbf{PRF} has been separately studied for the single-vector \cite{Dense-PRF} and multi-vector representation models \cite{Multi-PRF}.
    They exploit document- and term-level knowledge for query enrichment, respectively.
    We apply the corresponding PRF method for each backbone~model.
\end{itemize}% \vspace{-\topsep}
The second group uses inter-document similarity via a corpus~graph.
\begin{itemize}[leftmargin=*]% \vspace{-\topsep}
    \item \textbf{GAR} \cite{GAR} uses nearest neighbors in the graph to refine the initial ranking results. It is proposed for the reranking, but we apply it to the retrieval as well, as it brings consistent improvements.
    \item \textbf{LADR} \cite{LADR} uses lexical retriever in conjunction with the corpus graph to gradually expand the search space for PLM-based retriever. We use LADR-adaptive with no time constraints. 
    As LADR controls the search space, we apply it solely to $\mathcal{D}$.
\end{itemize}% \vspace{-\topsep}
As discussed in \cref{sec:relatedwork}, using topic knowledge for PLM-based retrievers has not been studied well.
Following a recent approach that uses generative augmentation \cite{mao2021generation, mackie2023generative}, we devise a new baseline that leverages topic knowledge discovered by PLMs.
\begin{itemize}[leftmargin=*]\vspace{-\topsep}
    \item \textbf{\TQE} uses generative query augmentation \cite{mao2021generation, mackie2023generative}. 
    Given a query, we extract its topic using PLMs, and enrich it by adding the predicted topics using the same template to \proposed (Eq.\ref{eq:tmplt}).
    We use T0-3B with the prompt proposed in \cite{mackie2023generative} (Appendix \ref{A:detail}).
    % \footnote{For topic generation, } 
\end{itemize}% \vspace{-\topsep}

\smallsection{Findings}
Table \ref{tab:retrieval} presents the retrieval results.
First, we observe that retrieval from the filtered corpus via SSA ($\mathcal{D}^{SSA}$) consistently yields higher recalls than retrieval from the entire corpus ($\mathcal{D}$), which again shows the effectiveness of our topic-based SSA.
Second, methods that utilize auxiliary corpus knowledge consistently boost the retrieval performance.
While \QE and PRF excel in the product domain, corpus graph knowledge demonstrates superior effectiveness in the academic domain. 
Conversely, \TQE, which leverages topic knowledge extracted using PLMs, shows limited effectiveness and even degrades the performance (e.g., R@1000, ColBERT in the academic domain). 
We notice that \TQE often fails to generate contexts reflecting domain knowledge.
For example, in Table \ref{tab:example}(a), it generates topics like ``data ownership'' and ``prove'', while relevant they do not reveal the high-level contexts of the academic paper.
Lastly, CRM consistently shows the highest recall in all setups.
Based on the taxonomy, it can identify topic classes reflecting domain knowledge (e.g., ``cryptography'', ``computer security'').
The topical relatedness is incorporated with the semantic similarity, providing complementary knowledge to each other.
These observations collectively show the effectiveness of using corpus taxonomy for theme-specific retrieval.
% Its high effectiveness shows the benefits of using the corpus taxonomy and also supports the validity of our design.
% Its high effectiveness supports our premise of using the corpus taxonomy as auxiliary corpus knowledge.
% and the validity of our design to leverage this corpus-level knowledge.

% \usepackage{booktabs}
% \usepackage{multirow}
\begin{table*}[t]
\caption{Reranking performance comparison on the academic and product domain. Significant differences with the baseline (i.e., reranking without using \proposed framework) are marked with * (p-value < 0.05 in the one-sample t-test).}
% \footnotesize
\renewcommand{\arraystretch}{0.85}
\resizebox{1.\linewidth}{!}{
\begin{tabular}{@{}cclcccccccccc@{}}
\toprule
\multirow{2}{*}{} & \multirow{2}{*}{\begin{tabular}[c]{c}Candidate set\\generation\end{tabular}} & \multirow{2}{*}{Method} & \multicolumn{5}{c}{MiniLM-L-12} & \multicolumn{5}{c}{MonoT5-base} \\ \cmidrule(lr){4-8} \cmidrule(lr){9-13}
 &  &  & N@3 & N@10 & MAP@10 & R@10 & R@50 & N@3 & N@10 & MAP@10 & R@10 & R@50 \\ \midrule\midrule
\multirow{9}{*}{\makecell[c]{Academic\\domain}} & \multirow{4}{*}{Retriever} & no reranking & 0.1589 & 0.1652 & 0.0966 & 0.1726 & 0.3166 & 0.1589 & 0.1652 & 0.0966 & 0.1726 & 0.3166 \\
 &  & Reranker & 0.1695 & 0.1760 & 0.1030 & 0.1827 & 0.3347 & 0.1748 & 0.1835 & 0.1078 & 0.1936 & \textbf{0.3368} \\
 &  & w/ GAR & \textbf{0.1701} & \textbf{0.1767} & \textbf{0.1036} & \textbf{0.1841} & \textbf{0.3358} & \textbf{0.1752} & \textbf{0.1854} & \textbf{0.1093} & \textbf{0.1972} & \textbf{0.3368} \\ 
  &  & w/ \TQE & 0.1666&	0.1753	&0.1023	&0.1822	&0.3301&	0.1727	&0.1800&	0.1060&	0.1876&	0.3309 \\
  \cmidrule(lr){2-13}
 &  & no reranking & 0.1748 & 0.1838 & 0.1074 & 0.1949 & 0.3633 & 0.1748 & 0.1838 & 0.1074 & 0.1949 & 0.3633 \\
 & \multirow{3}{*}{\begin{tabular}[c]{c}Retriever \\w/ \\\proposed-SSA, CRM\end{tabular}} & Reranker & 0.1780 & 0.1852 & 0.1090 & 0.1953 & 0.3634 & 0.1758 & 0.1868 & 0.1090 & 0.1997 & 0.3663 \\
 &  & w/ GAR & 0.1784 & 0.1870 & 0.1101 & 0.1979 & 0.3633 & 0.1794 & 0.1900 & 0.1118 & 0.2013 & 0.3671 \\
 &  & w/ \TQE & 0.1719&	0.1829&	0.1072&	0.1918&	0.3632&	0.1752&	0.1854&	0.1087&	0.1962&	0.3634 \\
 &  & \cellcolor{gray!15}w/ \proposed-QEP &\cellcolor{gray!15} \textbf{0.1821*} &\cellcolor{gray!15} \textbf{0.1915*} &\cellcolor{gray!15} \textbf{0.1126*} &\cellcolor{gray!15} \textbf{0.2026*} &\cellcolor{gray!15} \textbf{0.3660*} &\cellcolor{gray!15} \textbf{0.1828} &\cellcolor{gray!15} \textbf{0.1930*} &\cellcolor{gray!15} \textbf{0.1137} &\cellcolor{gray!15} \textbf{0.2048} &\cellcolor{gray!15} \textbf{0.3732*} \\ \midrule
  \multirow{9}{*}{\makecell[c]{Product\\domain}} & \multirow{4}{*}{Retriever} & no reranking & 0.2917 & 0.2845 & 0.1592 & 0.2401 & 0.4425 & 0.2917 & 0.2845 & 0.1592 & 0.2401 & 0.4425 \\
 &  & Reranker & \textbf{0.2972} & 0.2937 & 0.1664 & 0.2513 & 0.4544 & \textbf{0.3317} & 0.3214 & 0.1883 & \textbf{0.2642} & 0.4965 \\
  &  & w/ GAR & \textbf{0.2972} & \textbf{0.2986} & \textbf{0.1697} & \textbf{0.2610} & \textbf{0.4741} & \textbf{0.3317} & \textbf{0.3217} & \textbf{0.1892} & \textbf{0.2642} & \textbf{0.5061} \\ 
 &  & w/ \TQE & 0.2965	&0.2952	&0.1658&	0.2522&	0.4621&	0.3205&	0.3043&	0.1720&	0.2621	&0.4967 \\
\cmidrule(lr){2-13}
 &  & no reranking & 0.3048 & 0.2856 & 0.1603 & 0.2459 & 0.4627 & 0.3048 & 0.2856 & 0.1603 & 0.2459 & 0.4627 \\
 &\multirow{3}{*}{\begin{tabular}[c]{c}Retriever\\w/ \\\proposed-SSA, CRM\end{tabular}} & Reranker & 0.2903 & 0.2942 & 0.1644 & 0.2601 & 0.4689 & 0.3317 & 0.3215 & 0.1889 & 0.2642 & 0.5048 \\
 &  & w/ GAR & 0.3047 & 0.3021 & 0.1714 & 0.2654 & 0.4759 & 0.3318 & 0.3253 & 0.1916 & 0.2699 & 0.5098 \\
  &  & w/ \TQE & 0.2978&	0.3039	&0.1734&	0.2608&	0.4785&	0.3301	&0.3185&	0.1887&	0.2626&	0.5021  \\
 &  & \cellcolor{gray!15}w/ \proposed-QEP &\cellcolor{gray!15} \textbf{0.3189*} &\cellcolor{gray!15} \textbf{0.3139*} &\cellcolor{gray!15} \textbf{0.1818*} &\cellcolor{gray!15} \textbf{0.2701*} &\cellcolor{gray!15} \textbf{0.4891*} &\cellcolor{gray!15} \textbf{0.3416} &\cellcolor{gray!15} \textbf{0.3304*} &\cellcolor{gray!15} \textbf{0.1921} &\cellcolor{gray!15} \textbf{0.2729*} &\cellcolor{gray!15} \textbf{0.5227*}\\ \bottomrule
\end{tabular}}
\label{tab:reranking}
% \vspace{-0.2cm}
\end{table*}

\subsubsection{\textbf{Reranking}}
We assess the effectiveness of QEP for the second-stage reranking. 
Following \cite{GAR, thakur2021beir}, we use two backbone models: MiniLM-L-12 \cite{wang2020minilm}, MonoT5-base \cite{monot5}.
We report results for reranking top-100 results from both (a) retriever and (b) SSA~\&~CRM.\footnote{We use the retrieval results of \CTR-MS (for the academic domain) and SPLADE++ (for the product domain), which show the highest recalls within top-100.}

\smallsection{Compared methods}
We use GAR \cite{GAR}, the state-of-the-art method proposed for the reranking stage, as our main competitor.
We also compare \TQE. Note that QEP and \TQE only differ in the way of generating contexts to enrich~queries.

\smallsection{Findings}
In Table \ref{tab:reranking}, similar to results in Table \ref{tab:retrieval}, \TQE shows limited performance.
For the query in Table \ref{tab:example}(b), it generates contexts of ``dye'', ``peroxide'', and ``ammonia'', failing to add new information.
The best performance is consistently achieved by using all three strategies of \proposed.
QEP differs from \TQE in that it identifies core phrases using both the topic class knowledge and top-ranked results, under the idea of relevance model \cite{lavrenko2017relevance}.
These processes are guided by the taxonomy reflecting user-interested aspects, which may not be effectively revealed from the corpus graph.
Based on the findings in \cref{sec:results}, we conclude that each strategy of \proposed effectively enhances retrieval in each stage (RQ1, RQ2) and also has great compatibility with PLM-based models (RQ3).

% these processes are built upon the topical taxonomy reflecting user-interested aspects, which may not be effectively captured by the corpus graph.
% Also, it leverages topical taxonomy reflecting user-interested aspects not effectively captured by the corpus graph built upon the semantic similarity of documents.
% todo?: discussing inference cost of \proposed and competitors.

\begin{table}[t]
\caption{Ablation results for class relevance learning of training phase.}
\small
\renewcommand{\arraystretch}{0.6}
\resizebox{0.7\linewidth}{!}{
\begin{tabular}{cccc}
\toprule
\multicolumn{1}{l}{} & R@100 & R@500 & R@1000 \\ \midrule \midrule
\proposed-SSA, CRM & \textbf{0.4432} & \textbf{0.6399} & \textbf{0.7268} \\ \cmidrule(lr){1-4}
  $\mathbf{y}^s$ only & 0.3809 & 0.6080 & 0.7131 \\ 
    CKD $\rightarrow$ Self-training & 0.4327 & 0.6242 & 0.7226 \\
    \cmidrule(lr){1-4}
  w/o \proposed & 0.3783 & 0.5498 & 0.6216 \\ \bottomrule
\end{tabular}}
\label{tab:ablation1}
\vspace{-0.2cm}
\end{table}

% \usepackage{multirow}

\begin{table}[t]
\caption{Ablation results for each strategy of inference phase.}
\renewcommand{\arraystretch}{0.8}
\resizebox{0.8\linewidth}{!}{
\begin{tabular}{ccccc}
\toprule
 \multicolumn{1}{l}{} & \multicolumn{1}{l}{} & R@2500 & R@5000 & R@10000 \\\midrule \midrule
 \multirow{3}{*}{\rot{SSA}} & \proposed-SSA & \textbf{0.8439} & \textbf{0.9053} & \textbf{0.9599} \\
 & Low-level focus  & 0.7832 & 0.8784 & 0.9505 \\
 & High-level focus & 0.5569 & 0.6964 & 0.8552 \\ \midrule
 & \multicolumn{1}{l}{} & R@100 & R@500 & R@1000 \\ \midrule \midrule 
 \multirow{4}{*}{\rot{CRM}} & \proposed-SSA, CRM & \textbf{0.4432} & \textbf{0.6399} & \textbf{0.7268} \\
 & Low-level focus & 0.3864 & 0.5739 & 0.6623 \\
 & High-level focus & 0.3361 & 0.4606 & 0.5087 \\
 & w/o \proposed & 0.3783 & 0.5498 & 0.6216 \\ \midrule 
 & \multicolumn{1}{l}{} & N@10 & R@10 & R@50 \\ \midrule \midrule 
 \multirow{3}{*}{\rot{QEP}} & \proposed-SSA, CRM, QEP & \textbf{0.1915} & \textbf{0.2026} & \textbf{0.3660} \\
 & w/o top-ranked docs. & 0.1759 & 0.1855 & 0.3549 \\
  & w/o \proposed & 0.1760 & 0.1827 & 0.3347\\ \bottomrule
\end{tabular}}
\label{tab:abl2}
\vspace{0.3cm}
\end{table}

\subsection{Study of \proposed (RQ4)}
\label{subsec:study}
We provide a detailed ablation and hyperparameter study of \proposed.
We report results with \CTR-MS (for retriever) and MiniLM-L-12 (for reranker) on academic~domain.

\subsubsection{\textbf{Ablation study}}
\label{A:subsec_sup}
Table \ref{tab:ablation1} and Table \ref{tab:abl2} present ablation results for the training and inference phase, respectively.

\noindent
\textbf{Training phase.}
We compare two ablations intended to verify the effectiveness of our silver label generation and collective topic knowledge distillation (CKD):
\textbf{(1) $\mathbf{y}^s$ only} solely uses the generated silver labels for the class relevance learning without CKD, and
\textbf{(2) CKD $\rightarrow$ Self-training} replaces CKD with the standard self-training. 
The self-training is a well-established semi-supervised learning technique used to achieve better generalization when the given labels are incomplete \cite{self-training}. 
The core difference between self-training and CKD is that self-training generates additional training signals using model prediction on individual data instances, whereas CKD uses collective knowledge of the averaged prediction on semantically similar documents.
% Note that the taxonomy is corpus-level knowledge and ground-truth labels for classification (i.e., topic class labels for each document) are not available.
% For this reason, we analyze the effectiveness of the ablations in terms of the retrieval~performance.

We observe that both silver labels and CKD play important roles in class relevance learning.
First, the class relevance estimator only trained with $\mathbf{y}^s$ consistently improves the retrieval performance.
This supports the effectiveness of our silver label generation strategy.
Also, we observe that exploring relevant but unlabeled classes is highly important.
Both self-training and CKD bring considerable improvements compared to using only the silver label.
However, replacing CKD with self-training consistently degrades the retrieval effectiveness.
We also find that self-training is rather unstable compared to using CKD.

\begin{table}[t]
\small
\renewcommand{\arraystretch}{0.8}
\caption{Retrieval performance with varying retaining percent $m$.}
\begin{tabular}{cccc}\toprule
$m\, (\%)$ & Recall@100 & Recall@500 & Recall@1000 \\ \midrule\midrule
5 & 0.4396 & 0.6353 & 0.7216 \\
10 & \textbf{0.4432} & 0.6399 & 0.7268 \\
15 & 0.4414 & \textbf{0.6402} & 0.7274 \\
20 & 0.4410 & 0.6401 & \textbf{0.7278}\\\bottomrule
\end{tabular}
\label{tab:hp_m}
\vspace{-0.3cm}
\end{table}

\begin{table}[t]
\small
\caption{Reranking performance with varying sizes of the number of core phrases $k$.}
\renewcommand{\arraystretch}{0.7}
\begin{tabular}{cccc}
\toprule
$k$ & NDCG@10 & Recall@10 & Recall@50 \\ \midrule\midrule
2 & 0.1779 & 0.1888 & 0.3550 \\
5 & 0.1915 & 0.2026 & \textbf{0.3660} \\
7 & \textbf{0.1927} & \textbf{0.2044} & 0.3633 \\
10 & 0.1866 & 0.1956 & 0.3632 \\ \cmidrule(lr){1-4}
w/o \proposed & 0.1760 & 0.1827 & 0.3347\\ \bottomrule
\end{tabular}
\label{tab:hp_k}
\vspace{0.3cm}
\end{table}

\noindent
\textbf{Inference phase.}
We compare ablations for SSA, CRM, and QEP.
For SSA and CRM, which utilize the estimated class relevance distributions, we compare two alternative design choices:
\textbf{(1) low-level focus} emphasizes relevance to the more specific, narrower classes found at the lowermost levels, specifically the lowest two levels.
\textbf{(2) high-level focus} is the opposite choice of the low-level focus. 
It targets more general, broader classes closer to the root node, specifically the top three levels.
For QEP, we compare an ablation that ablates the use of top-ranked documents in the core phrase identification, denoted as \textbf{(3) w/o top-ranked docs.}
Specifically, from the set of relevant class phrases $\{P_j | \hat{b}_{qj}=1\}$, we collect the top-$k$ core phrases $P^q$ that most frequently appear across the corpus.

We observe that both low-level and high-level focus results in suboptimal results, and considering both board and specific topics consistently leads to the best recalls.
In particular, ignoring low-level classes (i.e., high-level focus) more drastically degrades the retrieval accuracy.
Also, we observe that using the frequency information from top-ranked documents is indeed effective in finding proper contexts to enrich queries.
For example, for the query in Table \ref{tab:example}(a), \textbf{w/o top-ranked docs} identifies core phrases like ``key'', ``scale'', ``software'', ``management'', and ``system'', which are relevant but not closely related to the query compared to the phrases obtained from \proposed.
These results support the validity of our design choice that uses the overall topic class distributions for SSA and CRM, and our core phrase identification based on the relevance model for~QEP.

\subsubsection{Hyperparameter study}
\label{A:hpstudy}
We provide analyses to guide the hyperparameter selection of \proposed.
Specifically, we investigate the effects of two hyperparameters introduced by \proposed: (1) $m$ denotes the retaining percent of topic classes (\cref{subsec:method_inference}), and (2) $k$ denotes the number of core phrases used to enrich queries in QEP (\cref{subsubsec:method_qep}).
We report the results in the academic domain. Similar tendencies are observed in the product domain.
First, Table \ref{tab:hp_m} presents the effects of $m$ on retrieval.
We observe stable performance overall with $m$ around $5-20\%$. 
In this work, we set $m=10\%$.
Second, Table \ref{tab:hp_k} presents the effects of $k$ on reranking.
We observe the best performance is achieved with $k$ around $5-7$. In this work, we set~$k=5$.

\subsection{Exploratory Analysis (RQ5, RQ6)}
We now explore the impact of labeled data and the quality of taxonomy on the effectiveness of \proposed.
We report the results with \CTR-MS (for retriever) and MiniLM-L-12 (for reranker) on the product domain. 
% which has training~labels.

\subsubsection{\textbf{\proposed with labeled data}}
\label{subsec:label_results}
While this work focuses on scenarios with no labeled data from the target corpus, \proposed can use labeled data for its training process (Appendix \ref{subsec:label}).
Table \ref{tab:labeled} presents the performance of \proposed with varying amounts of labeled data.
L0/L1/L2/L3 denote the setups using 0/33/66/100\% of the available training labels, respectively.
We observe that the overall performance of the retriever and reranker is largely improved by fine-tuning with labeled data.
Next, we observe that \proposed effectively enhances the retrieval process in all setups.
In specific, SSA effectively narrows down the initial search space without hurting recalls of the fine-tuned retriever.
Furthermore, CRM and QEP consistently improve both the retriever and reranker. 
These results show that \proposed can effectively leverage the labeled data, yielding a good synergy with the fine-tuned PLM-based models (RQ4).

% We believe high-quality synthetic data based on domain knowledge, can further improve retrieval in theme-specific applications.
% We note that recent methods have achieved promising results \cite{dai2022promptagator, chaudhary2023exploring} and explore the impacts of these synthetic data~in~future~studies.

\subsubsection{\textbf{Impacts of taxonomy quality}}
The power of \proposed is primarily attributed to the topical taxonomy.
While the methodology for taxonomy completion has been extensively studied and well-established, it's crucial to assess the robustness of \proposed regarding the quality of the given taxonomy. 
To this end, we consider two aspects measuring taxonomy quality \cite{lee2022taxocom}:
(1) topic completeness, which assesses how fully the topic nodes cover the true topics, and  
(2) term coherence, which assesses the semantic relatedness of terms (or phrases) within a topic node.
To impair completeness, we apply \textit{random pruning}, which randomly removes a node and all its child nodes.\footnote{We apply pruning by controlling the ratio of removed nodes to the total nodes.}  
To impair coherence, we apply \textit{level-wise node shuffling}, which randomly swaps nodes at the same level.\footnote{The root node has level $0$. We set the shuffling ratio as 10\%.} 
Note that such random shuffling corresponds to an extreme scenario.

\begin{figure}[t]
\centering
\includegraphics[width=0.47\linewidth]{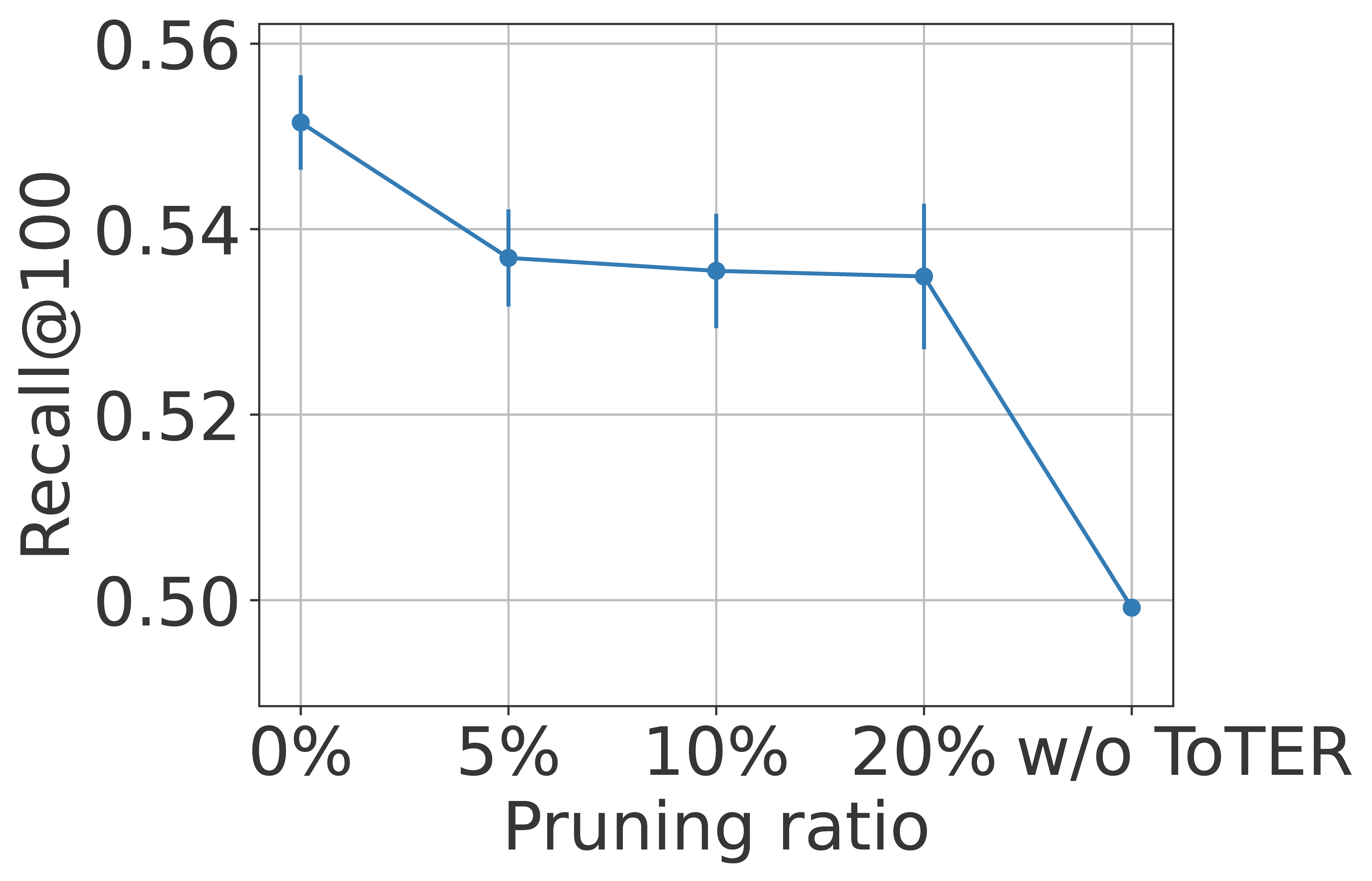}
\includegraphics[width=0.47\linewidth]{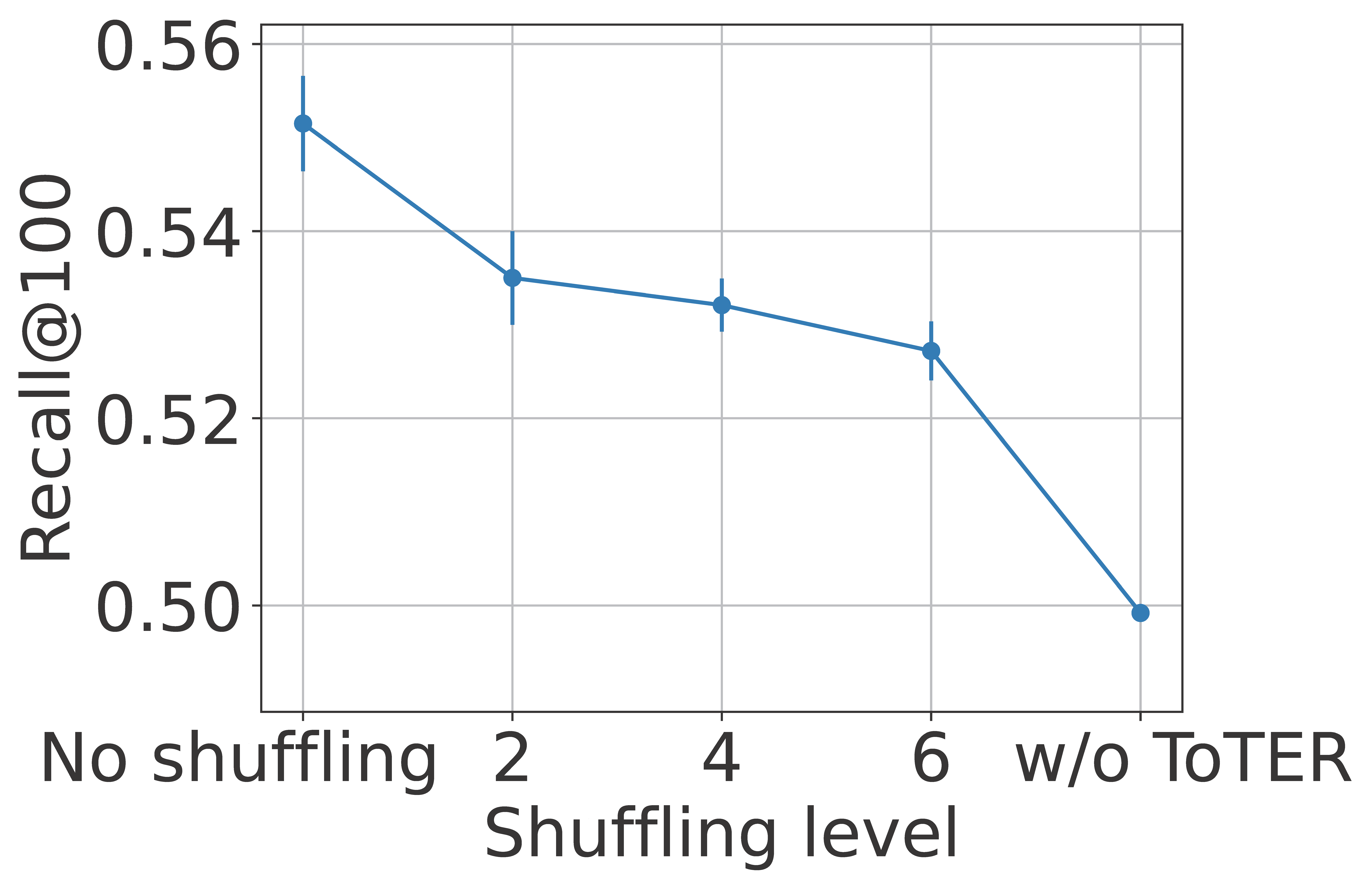}
% \caption*{(b) coherency}
\caption{Retrieval results with taxonomy impaired in terms of (left) topic completeness and (right) term coherence.}
\label{fig:robust}
\end{figure}

In Figure \ref{fig:robust}, we observe that although both types of noise degrade the effectiveness, \proposed has a considerable degree of robustness.
In particular, it shows a more stable performance for the pruning.
We conjecture this stability arises because the relevance to the removed nodes can be partially inferred from the relevance to the remaining topic nodes.
For the node shuffling, noise at deeper levels has higher impacts. 
This can be due to the increased node numbers at deeper levels and the decreased number of child nodes which can help to reduce the impacts of shuffling.
Based on the observations, we conclude that \proposed has a certain degree of robustness to taxonomy quality and can effectively enhance retrieval using the existing taxonomy completion methods (RQ5).

\begin{table}[t]
\caption{Performance of \proposed with labeled data. Significant differences with the baseline are marked with * (p-value < 0.05 in the one-sample (L0) /paired (L1-L3) t-test).}
\renewcommand{\arraystretch}{0.75}
\resizebox{0.98\linewidth}{!}{
\begin{tabular}{@{}clcccc@{}}
\toprule
 & Method & L0 & L1 & L2 & L3 \\ \midrule\midrule
\multirow{3}{*}{\rot{\small{R@100}}} & Retriever & 0.4992 & 0.5168 & 0.5179 & 0.5527 \\
 & w/ \proposed-SSA & 0.5009 & 0.5191 & 0.5202 & 0.5665 \\
 & w/ \proposed-SSA, CRM & \textbf{0.5515*} & \textbf{0.5502*} & \textbf{0.5865*} & \textbf{0.6098*} \\ \midrule
\multirow{3}{*}{\rot{\small{R@500}}} & Retriever & 0.6962 & 0.7592 & 0.7801 & 0.8071 \\ 
 & w/ \proposed-SSA & 0.7085 & 0.7697 & 0.7802 & 0.8090 \\
 & w/ \proposed-SSA, CRM & \textbf{0.7899*} & \textbf{0.7823*} & \textbf{0.8179*} & \textbf{0.8434*} \\ \midrule
\multirow{3}{*}{\rot{\small{R@1K}}} & Retriever & 0.8294 & 0.8333 & 0.8442 & 0.8714 \\ 
 & w/ \proposed-SSA & 0.8555 & 0.8353 & 0.8468 & 0.8718 \\
 & w/ \proposed-SSA, CRM & \textbf{0.8692*} & \textbf{0.8636*} & \textbf{0.8671*} & \textbf{0.8906*}\\ \bottomrule
\end{tabular}}
\begin{tabular}{@{}cccccc@{}} \end{tabular}
\resizebox{1.\linewidth}{!}{
\begin{tabular}{@{}cccccc@{}}
\toprule
L3 & N@3 & N@10 & MAP@10 & R@10 & R@50 \\ \midrule \midrule
Retriever \& Reranker & 0.3188 & 0.3107 & 0.1746 & 0.2616 & 0.5072 \\
w/ \proposed & \textbf{0.3241} & \textbf{0.3219} & \textbf{0.1907*} & \textbf{0.2778*} & \textbf{0.5249*}\\ \bottomrule
\end{tabular}}
\label{tab:labeled}
% \vspace{0.2cm}
\end{table}

\section{Conclusion}
\label{sec:conclusion}
We propose a new \proposed framework to enhance PLM-based retrieval in theme-specific applications using a corpus topical taxonomy. 
\proposed identifies the central topics of queries and documents with the guidance of topical taxonomy via class relevance learning, and exploits their topical relatedness to complement semantic matching by PLM-based models.
\proposed introduces three strategies, SSA, CRM, and QEP, which gradually focus on fine-grained ranking following the retrieve-then-rerank pipeline.
Our comprehensive experiments on two real-world datasets ascertain the benefits of using topical taxonomy and demonstrate the effectiveness of \proposed.
In future work, we plan to explore diverse approaches to leveraging the taxonomy in training retrievers and rerankers. 
% We believe that the taxonomy can enhance the existing techniques for synthetic query generation and identifier generation for generative retrieval.

\section*{Acknowledgements}
This work was supported IITP grant funded by MSIT (No.2018-0-00584, No.2019-0-01906), NRF grant funded by the MSIT (No.RS-2023-00217286, No.2020R1A2B5B03097210).
It was also in part by US DARPA KAIROS Program No. FA8750-19-2-1004 and INCAS Program No. HR001121C0165, National Science Foundation IIS-19-56151, and the Molecule Maker Lab Institute: An AI Research Institutes program supported by NSF under Award No. 2019897, and the Institute for Geospatial Understanding through an Integrative Discovery Environment (I-GUIDE) by NSF under Award No. 2118329. 

% Any opinions, findings, and conclusions or recommendations expressed herein are those of the authors and do not necessarily represent the views, either expressed or implied, of DARPA or the U.S. Government.

\pagebreak
\newpage
\clearpage

\bibliographystyle{ACM-Reference-Format} 
\balance
\bibliography{acmart}


\begin{thebibliography}{60}

%%% the Web address.

\ifx \showCODEN    \undefined \def \showCODEN     #1{\unskip}     \fi
\ifx \showDOI      \undefined \def \showDOI       #1{#1}\fi
\ifx \showISBNx    \undefined \def \showISBNx     #1{\unskip}     \fi
\ifx \showISBNxiii \undefined \def \showISBNxiii  #1{\unskip}     \fi
\ifx \showISSN     \undefined \def \showISSN      #1{\unskip}     \fi
\ifx \showLCCN     \undefined \def \showLCCN      #1{\unskip}     \fi
\ifx \shownote     \undefined \def \shownote      #1{#1}          \fi
\ifx \showarticletitle \undefined \def \showarticletitle #1{#1}   \fi
\ifx \showURL      \undefined \def \showURL       {\relax}        \fi
% invisible to TeX
\providecommand\bibfield[2]{#2}
\providecommand\bibinfo[2]{#2}
\providecommand\natexlab[1]{#1}
\providecommand\showeprint[2][]{arXiv:#2}

\bibitem[Arous et~al\mbox{.}(2023)]%
        {arous2023taxocomplete}
\bibfield{author}{\bibinfo{person}{Ines Arous}, \bibinfo{person}{Ljiljana Dolamic}, {and} \bibinfo{person}{Philippe Cudr{\'e}-Mauroux}.} \bibinfo{year}{2023}\natexlab{}.
\newblock \showarticletitle{TaxoComplete: Self-Supervised Taxonomy Completion Leveraging Position-Enhanced Semantic Matching}. In \bibinfo{booktitle}{\emph{WWW}}. \bibinfo{pages}{2509--2518}.
\newblock


\bibitem[Beltagy et~al\mbox{.}(2019)]%
        {SCIBERT}
\bibfield{author}{\bibinfo{person}{Iz Beltagy}, \bibinfo{person}{Kyle Lo}, {and} \bibinfo{person}{Arman Cohan}.} \bibinfo{year}{2019}\natexlab{}.
\newblock \showarticletitle{SciBERT: Pretrained Language Model for Scientific Text}. In \bibinfo{booktitle}{\emph{EMNLP}}.
\newblock
\showeprint{arXiv:1903.10676}


\bibitem[Blei et~al\mbox{.}(2003)]%
        {LDA}
\bibfield{author}{\bibinfo{person}{David~M Blei}, \bibinfo{person}{Andrew~Y Ng}, {and} \bibinfo{person}{Michael~I Jordan}.} \bibinfo{year}{2003}\natexlab{}.
\newblock \showarticletitle{Latent dirichlet allocation}.
\newblock \bibinfo{journal}{\emph{Journal of machine Learning research}} \bibinfo{volume}{3}, \bibinfo{number}{Jan} (\bibinfo{year}{2003}), \bibinfo{pages}{993--1022}.
\newblock


\bibitem[Chaudhary et~al\mbox{.}(2023)]%
        {chaudhary2023exploring}
\bibfield{author}{\bibinfo{person}{Aditi Chaudhary}, \bibinfo{person}{Karthik Raman}, \bibinfo{person}{Krishna Srinivasan}, \bibinfo{person}{Kazuma Hashimoto}, \bibinfo{person}{Mike Bendersky}, {and} \bibinfo{person}{Marc Najork}.} \bibinfo{year}{2023}\natexlab{}.
\newblock \showarticletitle{Exploring the Viability of Synthetic Query Generation for Relevance Prediction}.
\newblock \bibinfo{journal}{\emph{arXiv preprint arXiv:2305.11944}} (\bibinfo{year}{2023}).
\newblock


\bibitem[Cohan et~al\mbox{.}(2020)]%
        {SPECTER}
\bibfield{author}{\bibinfo{person}{Arman Cohan}, \bibinfo{person}{Sergey Feldman}, \bibinfo{person}{Iz Beltagy}, \bibinfo{person}{Doug Downey}, {and} \bibinfo{person}{Daniel~S. Weld}.} \bibinfo{year}{2020}\natexlab{}.
\newblock \showarticletitle{{SPECTER: Document-level Representation Learning using Citation-informed Transformers}}. In \bibinfo{booktitle}{\emph{ACL}}.
\newblock


\bibitem[Dai et~al\mbox{.}(2023)]%
        {dai2022promptagator}
\bibfield{author}{\bibinfo{person}{Zhuyun Dai}, \bibinfo{person}{Vincent~Y Zhao}, \bibinfo{person}{Ji Ma}, \bibinfo{person}{Yi Luan}, \bibinfo{person}{Jianmo Ni}, \bibinfo{person}{Jing Lu}, \bibinfo{person}{Anton Bakalov}, \bibinfo{person}{Kelvin Guu}, \bibinfo{person}{Keith~B Hall}, {and} \bibinfo{person}{Ming-Wei Chang}.} \bibinfo{year}{2023}\natexlab{}.
\newblock \showarticletitle{Promptagator: Few-shot dense retrieval from 8 examples}. In \bibinfo{booktitle}{\emph{ICLR}}.
\newblock


\bibitem[Devlin et~al\mbox{.}(2019)]%
        {BERT}
\bibfield{author}{\bibinfo{person}{Jacob Devlin}, \bibinfo{person}{Ming-Wei Chang}, \bibinfo{person}{Kenton Lee}, {and} \bibinfo{person}{Kristina Toutanova}.} \bibinfo{year}{2019}\natexlab{}.
\newblock \showarticletitle{{BERT}: Pre-training of Deep Bidirectional Transformers for Language Understanding}. In \bibinfo{booktitle}{\emph{NAACL-HLT}}.
\newblock


\bibitem[Dong et~al\mbox{.}(2022)]%
        {dong2022incorporating}
\bibfield{author}{\bibinfo{person}{Qian Dong}, \bibinfo{person}{Yiding Liu}, \bibinfo{person}{Suqi Cheng}, \bibinfo{person}{Shuaiqiang Wang}, \bibinfo{person}{Zhicong Cheng}, \bibinfo{person}{Shuzi Niu}, {and} \bibinfo{person}{Dawei Yin}.} \bibinfo{year}{2022}\natexlab{}.
\newblock \showarticletitle{Incorporating explicit knowledge in pre-trained language models for passage re-ranking}. In \bibinfo{booktitle}{\emph{SIGIR}}. \bibinfo{pages}{1490--1501}.
\newblock


\bibitem[Formal et~al\mbox{.}(2022)]%
        {SPLADE++}
\bibfield{author}{\bibinfo{person}{Thibault Formal}, \bibinfo{person}{Carlos Lassance}, \bibinfo{person}{Benjamin Piwowarski}, {and} \bibinfo{person}{St{\'e}phane Clinchant}.} \bibinfo{year}{2022}\natexlab{}.
\newblock \showarticletitle{From distillation to hard negative sampling: Making sparse neural ir models more effective}. In \bibinfo{booktitle}{\emph{SIGIR}}. \bibinfo{pages}{2353--2359}.
\newblock


\bibitem[Formal et~al\mbox{.}(2021)]%
        {SPLADE}
\bibfield{author}{\bibinfo{person}{Thibault Formal}, \bibinfo{person}{Benjamin Piwowarski}, {and} \bibinfo{person}{St{\'e}phane Clinchant}.} \bibinfo{year}{2021}\natexlab{}.
\newblock \showarticletitle{SPLADE: Sparse lexical and expansion model for first stage ranking}. In \bibinfo{booktitle}{\emph{Proceedings of the 44th International ACM SIGIR Conference on Research and Development in Information Retrieval}}. \bibinfo{pages}{2288--2292}.
\newblock


\bibitem[Gao and Callan(2022)]%
        {condenser}
\bibfield{author}{\bibinfo{person}{Luyu Gao} {and} \bibinfo{person}{Jamie Callan}.} \bibinfo{year}{2022}\natexlab{}.
\newblock \showarticletitle{Unsupervised Corpus Aware Language Model Pre-training for Dense Passage Retrieval}. In \bibinfo{booktitle}{\emph{ACL}}. \bibinfo{pages}{2843--2853}.
\newblock


\bibitem[Gao et~al\mbox{.}(2021)]%
        {simcse}
\bibfield{author}{\bibinfo{person}{Tianyu Gao}, \bibinfo{person}{Xingcheng Yao}, {and} \bibinfo{person}{Danqi Chen}.} \bibinfo{year}{2021}\natexlab{}.
\newblock \showarticletitle{{S}im{CSE}: Simple Contrastive Learning of Sentence Embeddings}. In \bibinfo{booktitle}{\emph{EMNLP}}. \bibinfo{pages}{6894--6910}.
\newblock


\bibitem[Hofst{\"a}tter et~al\mbox{.}(2021)]%
        {TASB}
\bibfield{author}{\bibinfo{person}{Sebastian Hofst{\"a}tter}, \bibinfo{person}{Sheng-Chieh Lin}, \bibinfo{person}{Jheng-Hong Yang}, \bibinfo{person}{Jimmy Lin}, {and} \bibinfo{person}{Allan Hanbury}.} \bibinfo{year}{2021}\natexlab{}.
\newblock \showarticletitle{Efficiently teaching an effective dense retriever with balanced topic aware sampling}. In \bibinfo{booktitle}{\emph{SIGIR}}. \bibinfo{pages}{113--122}.
\newblock


\bibitem[Huang et~al\mbox{.}(2020)]%
        {huang2020corel}
\bibfield{author}{\bibinfo{person}{Jiaxin Huang}, \bibinfo{person}{Yiqing Xie}, \bibinfo{person}{Yu Meng}, \bibinfo{person}{Yunyi Zhang}, {and} \bibinfo{person}{Jiawei Han}.} \bibinfo{year}{2020}\natexlab{}.
\newblock \showarticletitle{Corel: Seed-guided topical taxonomy construction by concept learning and relation transferring}. In \bibinfo{booktitle}{\emph{KDD}}. \bibinfo{pages}{1928--1936}.
\newblock


\bibitem[Iscen et~al\mbox{.}(2019)]%
        {pseudo-labeling}
\bibfield{author}{\bibinfo{person}{Ahmet Iscen}, \bibinfo{person}{Giorgos Tolias}, \bibinfo{person}{Yannis Avrithis}, {and} \bibinfo{person}{Ondrej Chum}.} \bibinfo{year}{2019}\natexlab{}.
\newblock \showarticletitle{Label propagation for deep semi-supervised learning}. In \bibinfo{booktitle}{\emph{CVPR}}. \bibinfo{pages}{5070--5079}.
\newblock


\bibitem[Izacard et~al\mbox{.}(2021)]%
        {CTR}
\bibfield{author}{\bibinfo{person}{Gautier Izacard}, \bibinfo{person}{Mathilde Caron}, \bibinfo{person}{Lucas Hosseini}, \bibinfo{person}{Sebastian Riedel}, \bibinfo{person}{Piotr Bojanowski}, \bibinfo{person}{Armand Joulin}, {and} \bibinfo{person}{Edouard Grave}.} \bibinfo{year}{2021}\natexlab{}.
\newblock \showarticletitle{Unsupervised dense information retrieval with contrastive learning}.
\newblock \bibinfo{journal}{\emph{arXiv preprint arXiv:2112.09118}} (\bibinfo{year}{2021}).
\newblock


\bibitem[Jian et~al\mbox{.}(2016)]%
        {topic_IR2}
\bibfield{author}{\bibinfo{person}{Fanghong Jian}, \bibinfo{person}{Jimmy~Xiangji Huang}, \bibinfo{person}{Jiashu Zhao}, \bibinfo{person}{Tingting He}, {and} \bibinfo{person}{Po Hu}.} \bibinfo{year}{2016}\natexlab{}.
\newblock \showarticletitle{A simple enhancement for ad-hoc information retrieval via topic modelling}. In \bibinfo{booktitle}{\emph{SIGIR}}. \bibinfo{pages}{733--736}.
\newblock


\bibitem[Johnson et~al\mbox{.}(2019)]%
        {johnson2019billion}
\bibfield{author}{\bibinfo{person}{Jeff Johnson}, \bibinfo{person}{Matthijs Douze}, {and} \bibinfo{person}{Herv{\'e} J{\'e}gou}.} \bibinfo{year}{2019}\natexlab{}.
\newblock \showarticletitle{Billion-scale similarity search with gpus}.
\newblock \bibinfo{journal}{\emph{IEEE Transactions on Big Data}} \bibinfo{volume}{7}, \bibinfo{number}{3} (\bibinfo{year}{2019}), \bibinfo{pages}{535--547}.
\newblock


\bibitem[Karpukhin et~al\mbox{.}(2020)]%
        {DPR}
\bibfield{author}{\bibinfo{person}{Vladimir Karpukhin}, \bibinfo{person}{Barlas Oguz}, \bibinfo{person}{Sewon Min}, \bibinfo{person}{Patrick Lewis}, \bibinfo{person}{Ledell Wu}, \bibinfo{person}{Sergey Edunov}, \bibinfo{person}{Danqi Chen}, {and} \bibinfo{person}{Wen-tau Yih}.} \bibinfo{year}{2020}\natexlab{}.
\newblock \showarticletitle{Dense Passage Retrieval for Open-Domain Question Answering}. In \bibinfo{booktitle}{\emph{EMNLP}}. \bibinfo{pages}{6769--6781}.
\newblock


\bibitem[Khattab and Zaharia(2020)]%
        {colbert}
\bibfield{author}{\bibinfo{person}{Omar Khattab} {and} \bibinfo{person}{Matei Zaharia}.} \bibinfo{year}{2020}\natexlab{}.
\newblock \showarticletitle{Colbert: Efficient and effective passage search via contextualized late interaction over bert}. In \bibinfo{booktitle}{\emph{SIGIR}}. \bibinfo{pages}{39--48}.
\newblock


\bibitem[Kim et~al\mbox{.}(2021)]%
        {kim2021query}
\bibfield{author}{\bibinfo{person}{Bosung Kim}, \bibinfo{person}{Hyewon Choi}, \bibinfo{person}{Haeun Yu}, {and} \bibinfo{person}{Youngjoong Ko}.} \bibinfo{year}{2021}\natexlab{}.
\newblock \showarticletitle{Query reformulation for descriptive queries of jargon words using a knowledge graph based on a dictionary}. In \bibinfo{booktitle}{\emph{CIKM}}. \bibinfo{pages}{854--862}.
\newblock


\bibitem[Kim et~al\mbox{.}(2022)]%
        {kim2022collective}
\bibfield{author}{\bibinfo{person}{Jihyuk Kim}, \bibinfo{person}{Minsoo Kim}, {and} \bibinfo{person}{Seung-won Hwang}.} \bibinfo{year}{2022}\natexlab{}.
\newblock \showarticletitle{Collective Relevance Labeling for Passage Retrieval}. In \bibinfo{booktitle}{\emph{NAACL-HLT}}. \bibinfo{pages}{4141--4147}.
\newblock


\bibitem[Kipf and Welling(2016)]%
        {GCN}
\bibfield{author}{\bibinfo{person}{Thomas~N Kipf} {and} \bibinfo{person}{Max Welling}.} \bibinfo{year}{2016}\natexlab{}.
\newblock \showarticletitle{Semi-supervised classification with graph convolutional networks}.
\newblock \bibinfo{journal}{\emph{arXiv preprint arXiv:1609.02907}} (\bibinfo{year}{2016}).
\newblock


\bibitem[Kulkarni et~al\mbox{.}(2023)]%
        {LADR}
\bibfield{author}{\bibinfo{person}{Hrishikesh Kulkarni}, \bibinfo{person}{Sean MacAvaney}, \bibinfo{person}{Nazli Goharian}, {and} \bibinfo{person}{Ophir Frieder}.} \bibinfo{year}{2023}\natexlab{}.
\newblock \showarticletitle{Lexically-Accelerated Dense Retrieval}. In \bibinfo{booktitle}{\emph{SIGIR}}. \bibinfo{pages}{152--162}.
\newblock


\bibitem[Lavrenko and Croft(2017)]%
        {lavrenko2017relevance}
\bibfield{author}{\bibinfo{person}{Victor Lavrenko} {and} \bibinfo{person}{W~Bruce Croft}.} \bibinfo{year}{2017}\natexlab{}.
\newblock \showarticletitle{Relevance-based language models}. In \bibinfo{booktitle}{\emph{ACM SIGIR Forum}}, Vol.~\bibinfo{volume}{51}. ACM New York, NY, USA, \bibinfo{pages}{260--267}.
\newblock


\bibitem[Lee et~al\mbox{.}(2022a)]%
        {lee2022taxocom}
\bibfield{author}{\bibinfo{person}{Dongha Lee}, \bibinfo{person}{Jiaming Shen}, \bibinfo{person}{SeongKu Kang}, \bibinfo{person}{Susik Yoon}, \bibinfo{person}{Jiawei Han}, {and} \bibinfo{person}{Hwanjo Yu}.} \bibinfo{year}{2022}\natexlab{a}.
\newblock \showarticletitle{Taxocom: Topic taxonomy completion with hierarchical discovery of novel topic clusters}. In \bibinfo{booktitle}{\emph{WWW}}. \bibinfo{pages}{2819--2829}.
\newblock


\bibitem[Lee et~al\mbox{.}(2022b)]%
        {lee2022topicgen}
\bibfield{author}{\bibinfo{person}{Dongha Lee}, \bibinfo{person}{Jiaming Shen}, \bibinfo{person}{Seonghyeon Lee}, \bibinfo{person}{Susik Yoon}, \bibinfo{person}{Hwanjo Yu}, {and} \bibinfo{person}{Jiawei Han}.} \bibinfo{year}{2022}\natexlab{b}.
\newblock \showarticletitle{Topic Taxonomy Expansion via Hierarchy-Aware Topic Phrase Generation}. In \bibinfo{booktitle}{\emph{Findings of the Association for Computational Linguistics: EMNLP 2022}}. \bibinfo{pages}{1687--1700}.
\newblock


\bibitem[Li et~al\mbox{.}(2023)]%
        {li2023sailer}
\bibfield{author}{\bibinfo{person}{Haitao Li}, \bibinfo{person}{Qingyao Ai}, \bibinfo{person}{Jia Chen}, \bibinfo{person}{Qian Dong}, \bibinfo{person}{Yueyue Wu}, \bibinfo{person}{Yiqun Liu}, \bibinfo{person}{Chong Chen}, {and} \bibinfo{person}{Qi Tian}.} \bibinfo{year}{2023}\natexlab{}.
\newblock \showarticletitle{SAILER: Structure-aware Pre-trained Language Model for Legal Case Retrieval}. In \bibinfo{booktitle}{\emph{SIGIR}}.
\newblock


\bibitem[Li et~al\mbox{.}(2021)]%
        {li2021topic}
\bibfield{author}{\bibinfo{person}{Xiangsheng Li}, \bibinfo{person}{Jiaxin Mao}, \bibinfo{person}{Weizhi Ma}, \bibinfo{person}{Yiqun Liu}, \bibinfo{person}{Min Zhang}, \bibinfo{person}{Shaoping Ma}, \bibinfo{person}{Zhaowei Wang}, {and} \bibinfo{person}{Xiuqiang He}.} \bibinfo{year}{2021}\natexlab{}.
\newblock \showarticletitle{Topic-enhanced knowledge-aware retrieval model for diverse relevance estimation}. In \bibinfo{booktitle}{\emph{WWW}}. \bibinfo{pages}{756--767}.
\newblock


\bibitem[Luan et~al\mbox{.}(2021)]%
        {luan2021sparse}
\bibfield{author}{\bibinfo{person}{Yi Luan}, \bibinfo{person}{Jacob Eisenstein}, \bibinfo{person}{Kristina Toutanova}, {and} \bibinfo{person}{Michael Collins}.} \bibinfo{year}{2021}\natexlab{}.
\newblock \showarticletitle{Sparse, dense, and attentional representations for text retrieval}.
\newblock \bibinfo{journal}{\emph{Transactions of the Association for Computational Linguistics}}  \bibinfo{volume}{9} (\bibinfo{year}{2021}), \bibinfo{pages}{329--345}.
\newblock


\bibitem[Ma et~al\mbox{.}(2020)]%
        {ma2020zero}
\bibfield{author}{\bibinfo{person}{Ji Ma}, \bibinfo{person}{Ivan Korotkov}, \bibinfo{person}{Yinfei Yang}, \bibinfo{person}{Keith Hall}, {and} \bibinfo{person}{Ryan McDonald}.} \bibinfo{year}{2020}\natexlab{}.
\newblock \showarticletitle{Zero-shot neural passage retrieval via domain-targeted synthetic question generation}.
\newblock \bibinfo{journal}{\emph{arXiv preprint arXiv:2004.14503}} (\bibinfo{year}{2020}).
\newblock


\bibitem[MacAvaney et~al\mbox{.}(2022)]%
        {GAR}
\bibfield{author}{\bibinfo{person}{Sean MacAvaney}, \bibinfo{person}{Nicola Tonellotto}, {and} \bibinfo{person}{Craig Macdonald}.} \bibinfo{year}{2022}\natexlab{}.
\newblock \showarticletitle{Adaptive re-ranking with a corpus graph}. In \bibinfo{booktitle}{\emph{CIKM}}. \bibinfo{pages}{1491--1500}.
\newblock


\bibitem[Mackie et~al\mbox{.}(2023)]%
        {mackie2023generative}
\bibfield{author}{\bibinfo{person}{Iain Mackie}, \bibinfo{person}{Shubham Chatterjee}, {and} \bibinfo{person}{Jeffrey Dalton}.} \bibinfo{year}{2023}\natexlab{}.
\newblock \showarticletitle{Generative Relevance Feedback with Large Language Models}. In \bibinfo{booktitle}{\emph{SIGIR}}.
\newblock


\bibitem[Mao et~al\mbox{.}(2021)]%
        {mao2021generation}
\bibfield{author}{\bibinfo{person}{Yuning Mao}, \bibinfo{person}{Pengcheng He}, \bibinfo{person}{Xiaodong Liu}, \bibinfo{person}{Yelong Shen}, \bibinfo{person}{Jianfeng Gao}, \bibinfo{person}{Jiawei Han}, {and} \bibinfo{person}{Weizhu Chen}.} \bibinfo{year}{2021}\natexlab{}.
\newblock \showarticletitle{Generation-Augmented Retrieval for Open-Domain Question Answering}. In \bibinfo{booktitle}{\emph{Proceedings of the 59th Annual Meeting of the Association for Computational Linguistics and the 11th International Joint Conference on Natural Language Processing (Volume 1: Long Papers)}}. \bibinfo{pages}{4089--4100}.
\newblock


\bibitem[Meng et~al\mbox{.}(2020)]%
        {meng2020hierarchical}
\bibfield{author}{\bibinfo{person}{Yu Meng}, \bibinfo{person}{Yunyi Zhang}, \bibinfo{person}{Jiaxin Huang}, \bibinfo{person}{Yu Zhang}, \bibinfo{person}{Chao Zhang}, {and} \bibinfo{person}{Jiawei Han}.} \bibinfo{year}{2020}\natexlab{}.
\newblock \showarticletitle{Hierarchical Topic Mining via Joint Spherical Tree and Text Embedding}. In \bibinfo{booktitle}{\emph{KDD}}.
\newblock


\bibitem[Nguyen et~al\mbox{.}(2016)]%
        {nguyen2016ms}
\bibfield{author}{\bibinfo{person}{Tri Nguyen}, \bibinfo{person}{Mir Rosenberg}, \bibinfo{person}{Xia Song}, \bibinfo{person}{Jianfeng Gao}, \bibinfo{person}{Saurabh Tiwary}, \bibinfo{person}{Rangan Majumder}, {and} \bibinfo{person}{Li Deng}.} \bibinfo{year}{2016}\natexlab{}.
\newblock \showarticletitle{Ms marco: A human-generated machine reading comprehension dataset}.
\newblock  (\bibinfo{year}{2016}).
\newblock


\bibitem[Nogueira et~al\mbox{.}(2020)]%
        {monot5}
\bibfield{author}{\bibinfo{person}{Rodrigo Nogueira}, \bibinfo{person}{Zhiying Jiang}, \bibinfo{person}{Ronak Pradeep}, {and} \bibinfo{person}{Jimmy Lin}.} \bibinfo{year}{2020}\natexlab{}.
\newblock \showarticletitle{Document Ranking with a Pretrained Sequence-to-Sequence Model}. In \bibinfo{booktitle}{\emph{Findings of the Association for Computational Linguistics: EMNLP 2020}}. \bibinfo{pages}{708--718}.
\newblock


\bibitem[Nogueira et~al\mbox{.}(2019)]%
        {DocT5query}
\bibfield{author}{\bibinfo{person}{Rodrigo Nogueira}, \bibinfo{person}{Jimmy Lin}, {and} \bibinfo{person}{AI Epistemic}.} \bibinfo{year}{2019}\natexlab{}.
\newblock \showarticletitle{From doc2query to docTTTTTquery}.
\newblock \bibinfo{journal}{\emph{Online preprint}}  \bibinfo{volume}{6} (\bibinfo{year}{2019}), \bibinfo{pages}{2}.
\newblock


\bibitem[Norouzi et~al\mbox{.}(2013)]%
        {binary_indexing}
\bibfield{author}{\bibinfo{person}{Mohammad Norouzi}, \bibinfo{person}{Ali Punjani}, {and} \bibinfo{person}{David~J Fleet}.} \bibinfo{year}{2013}\natexlab{}.
\newblock \showarticletitle{Fast exact search in hamming space with multi-index hashing}.
\newblock \bibinfo{journal}{\emph{IEEE transactions on pattern analysis and machine intelligence}} \bibinfo{volume}{36}, \bibinfo{number}{6} (\bibinfo{year}{2013}), \bibinfo{pages}{1107--1119}.
\newblock


\bibitem[Qu et~al\mbox{.}(2021)]%
        {rocketqa_v1}
\bibfield{author}{\bibinfo{person}{Yingqi Qu}, \bibinfo{person}{Yuchen Ding}, \bibinfo{person}{Jing Liu}, \bibinfo{person}{Kai Liu}, \bibinfo{person}{Ruiyang Ren}, \bibinfo{person}{Wayne~Xin Zhao}, \bibinfo{person}{Daxiang Dong}, \bibinfo{person}{Hua Wu}, {and} \bibinfo{person}{Haifeng Wang}.} \bibinfo{year}{2021}\natexlab{}.
\newblock \showarticletitle{{R}ocket{QA}: An Optimized Training Approach to Dense Passage Retrieval for Open-Domain Question Answering}. In \bibinfo{booktitle}{\emph{NAACL-HLT}}. \bibinfo{pages}{5835--5847}.
\newblock


\bibitem[Reddy et~al\mbox{.}(2022)]%
        {ESCI}
\bibfield{author}{\bibinfo{person}{Chandan~K Reddy}, \bibinfo{person}{Llu{\'\i}s M{\`a}rquez}, \bibinfo{person}{Fran Valero}, \bibinfo{person}{Nikhil Rao}, \bibinfo{person}{Hugo Zaragoza}, \bibinfo{person}{Sambaran Bandyopadhyay}, \bibinfo{person}{Arnab Biswas}, \bibinfo{person}{Anlu Xing}, {and} \bibinfo{person}{Karthik Subbian}.} \bibinfo{year}{2022}\natexlab{}.
\newblock \showarticletitle{Shopping queries dataset: A large-scale ESCI benchmark for improving product search}.
\newblock \bibinfo{journal}{\emph{arXiv preprint arXiv:2206.06588}} (\bibinfo{year}{2022}).
\newblock


\bibitem[Reimers and Gurevych(2019)]%
        {reimers-2019-sentence-bert}
\bibfield{author}{\bibinfo{person}{Nils Reimers} {and} \bibinfo{person}{Iryna Gurevych}.} \bibinfo{year}{2019}\natexlab{}.
\newblock \showarticletitle{Sentence-BERT: Sentence Embeddings using Siamese BERT-Networks}. In \bibinfo{booktitle}{\emph{EMNLP}}.
\newblock


\bibitem[Ren et~al\mbox{.}(2021)]%
        {rocketqa_v2}
\bibfield{author}{\bibinfo{person}{Ruiyang Ren}, \bibinfo{person}{Yingqi Qu}, \bibinfo{person}{Jing Liu}, \bibinfo{person}{Wayne~Xin Zhao}, \bibinfo{person}{QiaoQiao She}, \bibinfo{person}{Hua Wu}, \bibinfo{person}{Haifeng Wang}, {and} \bibinfo{person}{Ji-Rong Wen}.} \bibinfo{year}{2021}\natexlab{}.
\newblock \showarticletitle{{R}ocket{QA}v2: A Joint Training Method for Dense Passage Retrieval and Passage Re-ranking}. In \bibinfo{booktitle}{\emph{EMNLP}}. \bibinfo{pages}{2825--2835}.
\newblock


\bibitem[Robertson et~al\mbox{.}(2009)]%
        {BM25}
\bibfield{author}{\bibinfo{person}{Stephen Robertson}, \bibinfo{person}{Hugo Zaragoza}, {et~al\mbox{.}}} \bibinfo{year}{2009}\natexlab{}.
\newblock \showarticletitle{The probabilistic relevance framework: BM25 and beyond}.
\newblock \bibinfo{journal}{\emph{Foundations and Trends{\textregistered} in Information Retrieval}} \bibinfo{volume}{3}, \bibinfo{number}{4} (\bibinfo{year}{2009}), \bibinfo{pages}{333--389}.
\newblock


\bibitem[Shang et~al\mbox{.}(2020)]%
        {shang2020nettaxo}
\bibfield{author}{\bibinfo{person}{Jingbo Shang}, \bibinfo{person}{Xinyang Zhang}, \bibinfo{person}{Liyuan Liu}, \bibinfo{person}{Sha Li}, {and} \bibinfo{person}{Jiawei Han}.} \bibinfo{year}{2020}\natexlab{}.
\newblock \showarticletitle{Nettaxo: Automated topic taxonomy construction from text-rich network}. In \bibinfo{booktitle}{\emph{WWW}}. \bibinfo{pages}{1908--1919}.
\newblock


\bibitem[Shen et~al\mbox{.}(2018)]%
        {MAG_FS}
\bibfield{author}{\bibinfo{person}{Zhihong Shen}, \bibinfo{person}{Hao Ma}, {and} \bibinfo{person}{Kuansan Wang}.} \bibinfo{year}{2018}\natexlab{}.
\newblock \showarticletitle{A Web-scale system for scientific knowledge exploration}. In \bibinfo{booktitle}{\emph{Proceedings of ACL 2018, System Demonstrations}}. \bibinfo{pages}{87--92}.
\newblock


\bibitem[Thakur et~al\mbox{.}(2021)]%
        {thakur2021beir}
\bibfield{author}{\bibinfo{person}{Nandan Thakur}, \bibinfo{person}{Nils Reimers}, \bibinfo{person}{Andreas R{\"u}ckl{\'e}}, \bibinfo{person}{Abhishek Srivastava}, {and} \bibinfo{person}{Iryna Gurevych}.} \bibinfo{year}{2021}\natexlab{}.
\newblock \showarticletitle{{BEIR}: A Heterogeneous Benchmark for Zero-shot Evaluation of Information Retrieval Models}. In \bibinfo{booktitle}{\emph{Thirty-fifth Conference on Neural Information Processing Systems Datasets and Benchmarks Track (Round 2)}}.
\newblock


\bibitem[Wang and Zuccon(2023)]%
        {TASB-study}
\bibfield{author}{\bibinfo{person}{Shuai Wang} {and} \bibinfo{person}{Guido Zuccon}.} \bibinfo{year}{2023}\natexlab{}.
\newblock \showarticletitle{Balanced Topic Aware Sampling for Effective Dense Retriever: A Reproducibility Study}. In \bibinfo{booktitle}{\emph{SIGIR}}. \bibinfo{pages}{2542–2551}.
\newblock


\bibitem[Wang et~al\mbox{.}(2020)]%
        {wang2020minilm}
\bibfield{author}{\bibinfo{person}{Wenhui Wang}, \bibinfo{person}{Furu Wei}, \bibinfo{person}{Li Dong}, \bibinfo{person}{Hangbo Bao}, \bibinfo{person}{Nan Yang}, {and} \bibinfo{person}{Ming Zhou}.} \bibinfo{year}{2020}\natexlab{}.
\newblock \showarticletitle{Minilm: Deep self-attention distillation for task-agnostic compression of pre-trained transformers}.
\newblock \bibinfo{journal}{\emph{NeurIPS}}  \bibinfo{volume}{33} (\bibinfo{year}{2020}), \bibinfo{pages}{5776--5788}.
\newblock


\bibitem[Wang et~al\mbox{.}(2021)]%
        {Multi-PRF}
\bibfield{author}{\bibinfo{person}{Xiao Wang}, \bibinfo{person}{Craig Macdonald}, \bibinfo{person}{Nicola Tonellotto}, {and} \bibinfo{person}{Iadh Ounis}.} \bibinfo{year}{2021}\natexlab{}.
\newblock \showarticletitle{Pseudo-relevance feedback for multiple representation dense retrieval}. In \bibinfo{booktitle}{\emph{Proceedings of the 2021 ACM SIGIR International Conference on Theory of Information Retrieval}}. \bibinfo{pages}{297--306}.
\newblock


\bibitem[Wei and Croft(2006)]%
        {topic_IR1}
\bibfield{author}{\bibinfo{person}{Xing Wei} {and} \bibinfo{person}{W~Bruce Croft}.} \bibinfo{year}{2006}\natexlab{}.
\newblock \showarticletitle{LDA-based document models for ad-hoc retrieval}. In \bibinfo{booktitle}{\emph{SIGIRl}}. \bibinfo{pages}{178--185}.
\newblock


\bibitem[Xie et~al\mbox{.}(2016)]%
        {self-training}
\bibfield{author}{\bibinfo{person}{Junyuan Xie}, \bibinfo{person}{Ross Girshick}, {and} \bibinfo{person}{Ali Farhadi}.} \bibinfo{year}{2016}\natexlab{}.
\newblock \showarticletitle{Unsupervised deep embedding for clustering analysis}. In \bibinfo{booktitle}{\emph{ICML}}. PMLR, \bibinfo{pages}{478--487}.
\newblock


\bibitem[Xiong et~al\mbox{.}(2021)]%
        {ANCE}
\bibfield{author}{\bibinfo{person}{Lee Xiong}, \bibinfo{person}{Chenyan Xiong}, \bibinfo{person}{Ye Li}, \bibinfo{person}{Kwok-Fung Tang}, \bibinfo{person}{Jialin Liu}, \bibinfo{person}{Paul Bennett}, \bibinfo{person}{Junaid Ahmed}, {and} \bibinfo{person}{Arnold Overwijk}.} \bibinfo{year}{2021}\natexlab{}.
\newblock \showarticletitle{Approximate nearest neighbor negative contrastive learning for dense text retrieval}. In \bibinfo{booktitle}{\emph{ICLR}}.
\newblock


\bibitem[Yu et~al\mbox{.}(2021)]%
        {Dense-PRF}
\bibfield{author}{\bibinfo{person}{HongChien Yu}, \bibinfo{person}{Chenyan Xiong}, {and} \bibinfo{person}{Jamie Callan}.} \bibinfo{year}{2021}\natexlab{}.
\newblock \showarticletitle{Improving query representations for dense retrieval with pseudo relevance feedback}. In \bibinfo{booktitle}{\emph{CIKM}}. \bibinfo{pages}{3592--3596}.
\newblock


\bibitem[Zhan et~al\mbox{.}(2021)]%
        {zhan2021optimizing}
\bibfield{author}{\bibinfo{person}{Jingtao Zhan}, \bibinfo{person}{Jiaxin Mao}, \bibinfo{person}{Yiqun Liu}, \bibinfo{person}{Jiafeng Guo}, \bibinfo{person}{Min Zhang}, {and} \bibinfo{person}{Shaoping Ma}.} \bibinfo{year}{2021}\natexlab{}.
\newblock \showarticletitle{Optimizing dense retrieval model training with hard negatives}. In \bibinfo{booktitle}{\emph{SIGIR}}. \bibinfo{pages}{1503--1512}.
\newblock


\bibitem[Zhang et~al\mbox{.}(2018)]%
        {zhang2018taxogen}
\bibfield{author}{\bibinfo{person}{Chao Zhang}, \bibinfo{person}{Fangbo Tao}, \bibinfo{person}{Xiusi Chen}, \bibinfo{person}{Jiaming Shen}, \bibinfo{person}{Meng Jiang}, \bibinfo{person}{Brian Sadler}, \bibinfo{person}{Michelle Vanni}, {and} \bibinfo{person}{Jiawei Han}.} \bibinfo{year}{2018}\natexlab{}.
\newblock \showarticletitle{Taxogen: Unsupervised topic taxonomy construction by adaptive term embedding and clustering}. In \bibinfo{booktitle}{\emph{KDD}}. \bibinfo{pages}{2701--2709}.
\newblock


\bibitem[Zhang et~al\mbox{.}(2022a)]%
        {AR2}
\bibfield{author}{\bibinfo{person}{Hang Zhang}, \bibinfo{person}{Yeyun Gong}, \bibinfo{person}{Yelong Shen}, \bibinfo{person}{Jiancheng Lv}, \bibinfo{person}{Nan Duan}, {and} \bibinfo{person}{Weizhu Chen}.} \bibinfo{year}{2022}\natexlab{a}.
\newblock \showarticletitle{Adversarial Retriever-Ranker model for Dense Retrieval}. In \bibinfo{booktitle}{\emph{ICLR}}.
\newblock


\bibitem[Zhang et~al\mbox{.}(2022b)]%
        {zhang2022seed}
\bibfield{author}{\bibinfo{person}{Yu Zhang}, \bibinfo{person}{Yu Meng}, \bibinfo{person}{Xuan Wang}, \bibinfo{person}{Sheng Wang}, {and} \bibinfo{person}{Jiawei Han}.} \bibinfo{year}{2022}\natexlab{b}.
\newblock \showarticletitle{Seed-guided topic discovery with out-of-vocabulary seeds}. In \bibinfo{booktitle}{\emph{NAACL}}.
\newblock


\bibitem[Zhang et~al\mbox{.}(2023)]%
        {zhang2023effective}
\bibfield{author}{\bibinfo{person}{Yu Zhang}, \bibinfo{person}{Yunyi Zhang}, \bibinfo{person}{Martin Michalski}, \bibinfo{person}{Yucheng Jiang}, \bibinfo{person}{Yu Meng}, {and} \bibinfo{person}{Jiawei Han}.} \bibinfo{year}{2023}\natexlab{}.
\newblock \showarticletitle{Effective Seed-Guided Topic Discovery by Integrating Multiple Types of Contexts}. In \bibinfo{booktitle}{\emph{WSDM}}. \bibinfo{pages}{429--437}.
\newblock


\bibitem[Zheng et~al\mbox{.}(2020)]%
        {BERT-QE}
\bibfield{author}{\bibinfo{person}{Zhi Zheng}, \bibinfo{person}{Kai Hui}, \bibinfo{person}{Ben He}, \bibinfo{person}{Xianpei Han}, \bibinfo{person}{Le Sun}, {and} \bibinfo{person}{Andrew Yates}.} \bibinfo{year}{2020}\natexlab{}.
\newblock \showarticletitle{BERT-QE: Contextualized Query Expansion for Document Re-ranking}. In \bibinfo{booktitle}{\emph{Findings of the Association for Computational Linguistics: EMNLP 2020}}. \bibinfo{pages}{4718--4728}.
\newblock


\end{thebibliography}

\pagebreak
\newpage
\clearpage
\label{sec:appendix}
\appendix
\nobalance
\section{Appendix}
The source code of \proposed is provided through the author’s GitHub repository.\footnote{\url{https://github.com/SeongKu-Kang/ToTER_WWW24}}

\subsection{\proposed with Labeled Data}
\label{subsec:label}
While \proposed mainly focuses on scenarios without available labeled data from the target corpus, it can also benefit from harnessing $(q,d)$ labels. 
First, \proposed can directly use $q$ for class relevance learning (Eq.\ref{eq:bce}) by treating it as an additional document.
By using the class labels of its relevant document $d$, we can reflect their relevance into the class estimator. 
Also, the enriched queries by \proposed can be directly used to enhance the fine-tuning of the retriever and reranker.
That is, we use $q^{QEP}$ instead of $q$ as model input, where the fine-tuning follows the standard contrastive learning \cite{DPR}.

\subsection{Experiment details}

\subsubsection{\textbf{Corpus Topical taxonomy construction}}
\label{A:taxoconstruction}
The corpus topical taxonomy is obtained by applying the existing taxonomy completion technique \cite{lee2022taxocom} on a seed taxonomy.
Based on the seed taxonomy, we conduct taxonomy completion which completes and adjusts the taxonomy for the target corpus.
This is a critical step to ensure the taxonomy aligns with the target corpus.
Note that the seed taxonomy is incomplete;
it contains numerous topics irrelevant to the corpus as well as failing to cover all topics.
For example, in the case of the SCIDOCS dataset, we discovered that over 95\% of topic classes from the seed taxonomy have no (or a very weak) relevance to documents in the corpus.
Also, as it does not cover all specific topics in the corpus, we need to expand it by identifying new topics not present in the original seed taxonomy.
We use the recently proposed technique \cite{lee2022taxocom} to obtain the corpus topical taxonomy.
We use the official implementation provided by the authors.\footnote{\url{https://github.com/donalee/taxocom/tree/main}}
A notable change from the original implementation is that we additionally use PLM knowledge for more effective topic discovery \cite{zhang2023effective}.

\subsubsection{\textbf{Implementation details}}
\label{A:detail}
In our experiments, we use \texttt{BEIR} benchmark framework\footnote{\url{https://github.com/beir-cellar/beir}} for evaluating all compared methods.
For BM25, we use Elasticsearch.
For docT5query and GenQ, we use T5 models with checkpoints provided by \texttt{BEIR}.\footnote{\texttt{castorini/doc2query-t5-base-msmarco}, \texttt{BeIR/query-gen-msmarco-t5-base-v1}}
For all compared PLM-based retrieval and reranking models, we use checkpoints that are publicly available:
\CTR-MS\footnote{\texttt{facebook/contriever-msmarco}}, TAS-B\footnote{\texttt{msmarco-distilbert-base-tas-b}}, ColBERT\footnote{\url{https://github.com/terrierteam/pyterrier_colbert}}, SPLADE++\footnote{\texttt{naver/splade-cocondenser-ensembledistil}}, MiniLM-L-12\footnote{\texttt{cross-encoder/ms-marco-MiniLM-L-12-v2}}, and MonoT5-base\footnote{\texttt{castorini/monot5-base-msmarco-10k}}.
For multi-vector representation models, we use ColBERT.v1 solely for compatibility with the public ColBERT-PRF implementation.

To generate the corpus graph, we use \CTR-MS for the academic domain and SPLADE++ for the product domain, as they consistently show the highest recalls within top-100.
The number of neighbors in the graph is set as $10$, as it shows stable results in both GAR \cite{GAR} and LADR \cite{LADR}.
For BERT-QE \cite{BERT-QE}, we set both the number of top-ranked documents and the number of core segments (or chunks) as 10, following the paper.
For PRF of the single and sparse representation model (i.e., \CTR-MS, SPLADE++), we set the number of documents for query enrichment as $5$.
For PRF for the multi-vector representation model (i.e., ColBERT), we use the official implementation and provided values.
For other baseline-specific hyperparameters, we follow the recommended values in the original papers \cite{Dense-PRF, Multi-PRF, GAR, LADR}.
For \TQE, we use T0-3B\footnote{\texttt{bigscience/T0\_3B}} with the prompt suggested in \cite{mackie2023generative}: \textit{"Based on the query, generate a bullet-point list of relevant topics present in relevant documents:"}.

For fine-tuning with labeled data in the product domain (\cref{subsec:label_results}), we use \texttt{sentence\_transformers} framework \cite{reimers-2019-sentence-bert}.
We continue fine-tuning for $10$ epochs with a learning rate of $7e^{-5}$. 
We discovered that further increasing the training epochs consistently degrades the retrieval accuracy.
For \proposed, we set the retaining percent $m=10\%$, the number of core phrases $k=5$. 
%We provide a sensitivity study with varying $m$ and $k$ in \ref{A:hpstudy}.
% For the similarity function, we use the inner product of the ranking score.
Following BERT-QE \cite{BERT-QE}, the number of top-ranked documents (for collective labels and QEP) is set to $10$.
% Lastly, the number of GNN layers ($L$) in the class relevance estimator is set to $1$, but it can be further tuned.

\subsection{Supplementary Results}
\label{A:case_study}
We provide additional case studies of \proposed and generation-based baselines (i.e., GenQ, \TQE).
Table \ref{tab:case1} presents the generated contexts from GenQ and \TQE for examples in Table \ref{tab:example}.
We observe that the generated queries and topics do not effectively reveal the domain-specific contexts.
These results again show the difficulty of retrieval in theme-specific applications and the importance of proper use of well-structured corpus knowledge.

Lastly, Table \ref{tab:example2} presents additional case studies for each domain.
Similar to the previous case studies in Table \ref{tab:example}, \proposed can effectively identify common topic classes and core topic phrases, guided by the theme-specific taxonomy.
This information complements the semantic matching by PLM-based retrieval models, allowing the models to more accurately find relevant documents.

\begin{table}[h]
\large
\caption{Case studies of generation-based baselines for examples in Table \ref{tab:example}. For GenQ for the product domain, we present results of Document A.}
\resizebox{1.\linewidth}{!}{
\begin{tabular}{p{1.8cm}|p{3.5cm}|p{3.5cm}} \hlineB{2.5}
 & (a) Academic domain & (b) Product   domain \\ \hline\hline
\multirow{2}{*}{GenQ} & ``what are pors.'' & ``what is the name for the hair dyes.'' \\
 & ``what is por.'' & ``what is onc hair color.'' \\ \hline
\TQE & data ownership, prove, store, proving, possession & dye, peroxide, ammonia, hair, natural \\\hlineB{2.5}
\end{tabular}}
\label{tab:case1}
\end{table}

\begin{table*}[b]
\caption{Additional case studies of \proposed.
We use \CTR-MS (retriever) and MiniLM-L-12 (reranker). 
Contents closely related to the query are denoted in bold.}
\large
\renewcommand{\tabcolsep}{0.9mm}
\resizebox{1.\linewidth}{!}{
\begin{tabular}{p{1.8cm}|p{8.5cm}|p{2.04cm}|p{8.5cm}}
\hlineB{2.5}
\multicolumn{2}{c|}{(a) Academic domain} & \multicolumn{2}{c}{(b) Product domain} \\ \hline\hline
Query & Exploring venue popularity in foursquare & Query & 0.07 lash extensions not easy fans \\ \hline 
\multirow{1}{=}{Document A (label: relevant, \\rank:\\top-167)} 
& \multirow{1}{=}{\parbox{8.5cm}{\vspace{2pt} \textbf{Event-based social networks}: linking the \textbf{online and offline social worlds}. 
Newly emerged event-based online \textbf{social services}, such as Meetup and Plancast, have experienced increased \textbf{popularity} and rapid growth. From these services, we observed a new type of social network - event-based social network (EBSN). ... This paper is the first research to study EBSNs at scale and paves the way for future studies on this new type of social network.}}
& \multirow{1}{=}{Document A (label: relevant, \\rank: top-57)}
& Volume \textbf{Lash Extensions} \textbf{0.07} D Curl Mix 8-15mm Eyelash Extensions. ... One Second easily grafted \textbf{eyelash}, a special craft that allowing you to create 2D10D \textbf{fans}. Thickness 0.03/0.05/\textbf{0.07}/0.10mm. CurlC/CC/D/DD curl \textbf{lash extensions}. Single length 8mm25mm. Mixed length 8-15mm MIX, 9-16mm MIX, 15-20mm MIX in one tray. The root of the volume \textbf{lashes extensions} will not separate, any flowering and novices can operate. ... \\ \hlineB{2.5}
\multirow{2}{=}{\proposed rank:\\ top-28 }& Topic classes: diffusion, social network, social behavior,  world wide web,  computer science, economy, economics &  \multirow{2}{=}{\proposed rank: top-7}  & Topic classes: makeup brushes and tools, false lashes, tools and accessories, eye, beauty and personal care \\ \cline{2-2}\cline{4-4}
 & \multirow{1}{=}{Core phrases: point of interest,  location, check, social network, foursquare} &  & Core phrases: eyelash, fan, extensions, volume lash extensions, lash\\ 
\hlineB{2.5}
\end{tabular}}
\label{tab:example2}
\vspace{-0.25cm}
\end{table*}

% % \subsubsection{Impacts of taxonomy quality (R@1000)}
% % Figure \ref{fig:robust2} provides Recall@1000 results with impaired taxonomies.
% % Similar to Recall@100 results in Figure \ref{fig:robust}, \proposed shows a considerable degree of robustness.
% % Interestingly, 
% % \begin{figure}[t]
% % \centering
% % \includegraphics[width=0.47\linewidth]{images/ESCI_pruning1000.pdf}
% % \includegraphics[width=0.47\linewidth]{images/ESCI_shuffling1000.pdf}
% % \caption{Retrieval results with taxonomy impaired in terms of (left) topic completeness and (right) term coherence.}
% % \label{fig:robust2}
% % \end{figure}

\end{document}